\documentclass{emulateapj}
\usepackage{natbib}

\usepackage{amsmath,amssymb,amsthm,mathrsfs,dsfont}
\usepackage{graphicx}
\usepackage{subfigure}
\usepackage{float}
\usepackage{booktabs}
\usepackage{epstopdf}
\usepackage{color}

\def\be{\begin{equation}}
\def\ee{\end{equation}}

\def\kms{{\rm \,km\,s^{-1}}}

\def\Gyr{{\rm \,Gyr}}
\def\e{{\rm e}}
\def\Mpc{{\rm \,Mpc}}
\def\kpc{{\rm \,kpc}}

\def\keV{{\rm \,keV}}
\def\K{{\rm \,K}}
\def\msun{{\,M_\odot}}

\begin{document}

\title{OFFSETS BETWEEN THE X-RAY AND THE SUNYAEV-ZEL'DOVICH-EFFECT
PEAKS IN MERGING GALAXY CLUSTERS AND THEIR COSMOLOGICAL IMPLICATIONS}
\shortauthors{Zhang, Yu, \& Lu}

\author{Congyao Zhang$^1$, Qingjuan Yu$^1$, and Youjun Lu$^2$}
\affil{
$^1$~Kavli Institute for Astronomy and Astrophysics, Peking
University, Beijing, 100871, China; yuqj@pku.edu.cn \\
$^2$~National Astronomical Observatories, Chinese Academy of
Sciences, Beijing, 100012, China
}

\begin{abstract}

Observations reveal that the peaks of the X-ray map and the Sunyaev-Zel'dovich
(SZ) effect  map of some galaxy clusters are offset from each other. In this
paper, we perform a set of hydrodynamical simulations of mergers of two galaxy
clusters to investigate the spatial offset between the maxima of the X-ray and
the SZ surface brightness of the merging clusters. We find that significantly
large SZ-X-ray offsets ($>100 \kpc$) can be produced during the major mergers
of galaxy clusters (with mass $>1\times 10^{14}\msun$). The significantly large
offsets are mainly caused by a `jump effect' occurred between the primary and
secondary pericentric passages of the two merging clusters, during which the
X-ray peak may jump to the densest gas region located near the center of the
small cluster, but the SZ peak remains near the center of the large one.  Our
simulations show that merging systems with higher masses and larger initial
relative velocities may result in larger offset sizes and longer offset time
durations; and only nearly head-on mergers are likely to produce significantly
large offsets. We further investigate the statistical distribution of the
SZ-X-ray offset sizes and find that (1) the number distribution of the offset
sizes is bimodal with one peak located at low offsets $\sim 0$ and the other at
large offsets $\sim$350--450$ h^{-1}\kpc$, but the objects with intermediate
offsets are scarce; and (2) the probabilities of the clusters in the mass range
higher than $2\times10^{14} h^{-1}\msun$ that have offsets larger than 20, 50,
200, 300, and $500 h^{-1}\kpc$ are 34.0\%, 11.1\%, 8.0\%, 6.5\%, and 2.0\%
respectively at $z=0.7$. The probability is sensitive to the underlying
pairwise velocity distribution and the merger rate of clusters.  We suggest
that the SZ-X-ray offsets provide a probe to the cosmic velocity fields on the
cluster scale and the cluster merger rate, and future observations on the
SZ-X-ray offsets for a large number of clusters may put strong constraints on
them.

Our simulation results suggest that the SZ-X-ray offset in the Bullet Cluster,
together with the mass ratio of the two merging clusters, requires a relative
velocity larger than $3000\kms$ at an initial separation $5 \Mpc$.  The
cosmic velocity distribution at the high-velocity end is expected to be crucial
in determining whether there exists an incompatibility between the existence of
the Bullet Cluster and the prediction of a $\rm \Lambda CDM$ model.

\end{abstract}

\section{Introduction}

Clusters of galaxies, the largest virialized systems known in the universe, are
formed from mergers of small structures in the hierarchical structure formation
and evolution model \citep[see a review in][]{Kravtsov2012}. Collisions of
galaxy clusters, with relative velocities up to several hundred or several
thousand $\kms$, are probably the most energetic events since the big bang,
which re-distribute both the dark matter (DM) and the baryonic matter in the
colliding clusters at an Mpc scale and gas can be shocked and heated. A number
of observational features have indicated that some cluster systems are
undergoing mergers or are the remnants of recent mergers. For example, the
spatial separation of the X-ray emitting gas and the DM clump in the Bullet
Cluster is explained by a collision of two clusters, in which gas interacts
electromagnetically and moves slower than DM \citep{Clowe2004, Clowe2006}; and
the sharp edges shown in some cluster X-ray images are interpreted as either
the `shock fronts' or `cold fronts' formed in the intracluster medium (ICM)
during cluster collisions \citep[][]{Markevitch1999, Markevitch2000,MV07}. In
addition, recent Sunyaev-Zel'dovich (SZ) cluster surveys, including the South
Pole Telescope (SPT), the Atacama Cosmology Telescope (ACT), and the Planck SZ
surveys, show that the positions of the maxima of the X-ray surface brightness
and the SZ effect differ significantly in some clusters \citep{Andersson2011,
PlanckE8, Menanteau2012}; and also collisions of clusters can lead to the
position displacement \citep{PlanckE9}.  In this paper we study how the spatial
displacement of the X-ray and the SZ signals from a merging cluster is affected
by the merging process and explore how the statistical distribution of the
displacements is connected to the cluster pairwise velocity field and the
cluster merger rate. Study of the cluster merging processes and the possible
observational signatures would improve our understanding of the baryonic
physics involved in cluster collisions, the cosmic velocity fields, and further
the structure formation and evolution model and the underlying cosmological
model.

Combination of the X-ray properties and the SZ effects of clusters has been
used before to constrain the cosmological parameters \citep{Carlstrom2002} and
investigate baryonic physics in clusters of galaxies
\citep[e.g.,][]{Andersson2011,PlanckE10,PlanckM1}. The X-ray emission of the
ICM gas comes mainly from the thermal bremsstrahlung radiation of the hot
electrons, and the X-ray luminosity is proportional to $n_\e^2T^{1/2}_\e$,
where $n_\e$ is the number density of electrons and $T_\e$ is the electron
temperature.  The SZ effect is the result of energy boost of low-energy cosmic
microwave background photons due to the inverse Compton scattering off
high-energy electrons in the ICM.  Depending on whether the high energy of the
electrons is due to their temperature (thermal) or bulk motion (kinematic), the
SZ effect can be divided into the thermal effect and the kinematic effect. The
thermal SZ effect has been detected in many clusters, and its magnitude is
proportional to the integral of the electron pressure along the line of sight
($\propto n_\e T_\e$) and independent of redshift.  The kinematic effect is
expected to be a potential probe to the motion of clusters, yet it is much
weaker than the thermal effect in high-mass clusters. As the thermal SZ effect
and the X-ray brightness of a cluster have a different dependence on gas
temperature and density distribution, and the location of the maximum X-ray
brightness is likely to be offset from the location of the maximum of the SZ
effect because of the re-distribution of the gas in the cluster merging
process.

In this paper we use numerical simulations to study the merging processes of
individual clusters and then obtain the distribution maps of the X-ray
brightness and the SZ effect in the merging clusters. Many simulations on
mergers of two individual clusters have been performed previously to produce
the observed configuration/morphology of a specific unrelaxed galaxy cluster
\citep[e.g.,][]{Springel2007, Mastropietro2008, Zuhone2009} or investigate
various physical effects and features caused by the merging processes
\citep[see e.g.,][]{Roettiger1997, Ritchie2002, Poole2006, Zuhone2011}.
\citet{Molnar2012} simulated mergers of two clusters to particularly reconstruct
the morphology of the galaxy cluster CL 0152-1357 and the offset between its SZ
effect and X-ray peaks; and in their study, only the high relative velocity
($>3000\kms$) case of two clusters was explored, which is extremely rare in the
universe. In our study, we explore the processes in a larger range of the
parameter space (e.g., in the initial kinematic distribution of colliding
clusters, their total masses and mass ratios), so that the distribution of the
offsets can be investigated statistically.
We find that the distribution of the SZ-X-ray offset is mainly
affected by the merger rate of clusters and the pairwise velocity distribution
of those merging clusters. Therefore, observational estimates of the SZ-X-ray
offset distribution by SZ and X-ray surveys can be used to put constraints on
both the merger rate of clusters and the pairwise velocity distribution of
clusters, and further on the $\Lambda$CDM model at the cluster scale.

This paper is organized as follows. In Section 2, we describe our numerical
methods of the cluster mergers and the initial conditions of the simulations.
In Section 3, we present the general results of the simulations and discuss the
factors to affect the offsets between the SZ effect and the X-ray peaks. In
Section~\ref{sec:prob}, we estimate the probability of the large offsets and
compare it with observations.  Finally, conclusions are summarized in
Section~\ref{sec:conclusion}.

Throughout the paper, we use a flat $\rm \Lambda CDM$ cosmology model with
$\Omega_{\rm m}=0.30$, $\Omega_{\rm b}=0.05$, $\Omega_{\rm \Lambda}=0.70$,
$H=100h\kms\Mpc^{-1}$ with $h=0.7$.

\section{Method}

We perform numerical simulations of the mergers of two galaxy clusters. In this
work, each of the clusters is simplified as a halo with a mixture of only DM
and gas. The DM is approximated as collisionless, undergoing only gravity; and
the gaseous component is collisional, adiabatic, undergoing both gravity and
fluid pressure. Both the particle-based Lagrangian [such as smoothed particle
hydrodynamics (SPH)] method \citep{Ritchie2002, Poole2006} and the mesh-based
Eulerian [such as adaptive mesh refinement (AMR)] method \citep{Zuhone2011,
Molnar2012} have been used to carry out such simulations.  \citet{Mitchell2009}
investigated the discrepancy occurred in the merger simulations between these
two different methods, by using GADGET-2 (SPH; \citealt{Springel2001,
Springel2005}) and FLASH (AMR) codes. They showed that SPH has the advantages
in computational speed, effective resolution, and Galilean invariance, but is
not good at the treatment of eddies and fluid instabilities, while AMR is on
the contrary.  Considering the purpose of our work, we choose the SPH code to
do the simulations for the following reasons: (1) we need to explore a large
parameter space of the merging processes, which demands an efficient
calculation speed; (2) we focus on the peak positions of the X-ray and the SZ
signals, and the disadvantages of the SPH code do not have significant effects
on these; and (3) by applying both the AMR code and the SPH code to simulate
the mergers for some cases in Section~\ref{sec:result_sphamr}, and then
comparing both the simulation results concerned in this study, we find that
different numerical codes do not lead to significant changes.

We have made some simplifications in simulating the physical processes occurred
in the clusters. (1) We do not include radiative cooling and various heating
mechanisms in our simulations.  A pure cooling model has to face the
overcooling problem, which is inconsistent with observations
\citep{Suginohara1998}; and thus an effective heating form is necessitated to
counterbalance the cooling effect, such as AGN feedback \citep{Sijacki2007}.
However, the physics of those mechanisms is not well understood yet. As argued
in \citet{Zuhone2011}, the simplified model for gas physics can serve as a
baseline to characterize the effect that we are interested in this work.  In
addition, \citet{Mastropietro2008} showed little evidence that the radiative
cooling can change the positions of X-ray peaks, though the surface brightness
is obviously modified. (2) The magnetic field is omitted in our simulations,
and we assume that it has little influence on the SZ-X-ray offsets.  In our
work SZ-X-ray offset is formed from the breaking of hydrostatic equilibrium of
the ICM gas during cluster energetic collisions. The magnetic energy is usually
smaller than $1\%$ of the mechanical energy involved in cluster mergers
\citep{Carilli2002}, and on average the magnetic pressure in clusters is much
smaller than the thermal pressure \citep{Lagan2010}.  Moreover, the tangling
scale of the magnetic field is about $10\kpc$ \citep{Carilli2002}, which is an
order of magnitude smaller than the typical offsets discussed in this work.
Nevertheless, the quantitative analysis on the effects after including the
ignored physical processes here needs to be investigated in the future.

\subsection{Initial distributions of the dark matter and the gaseous halo of a
cluster}

We assume a spherical symmetric density profile to model the initial mass
distributions of the DM and the gas in a cluster.  The Navarro-Frenk-White
(NFW) profile \citep{Navarro1997} is used for the DM mass density distribution
within the virial radius $r_{\rm vir}$,
\begin{equation}
\rho_{\rm DM}(r) = \frac{\rho_{\rm s}}{r/r_{\rm s}(1+r/r_{\rm s})^{2}}, {\ \rm
if\ } r \leq r_{\rm vir},
\label{eq:rhoDMrltvir}
\end{equation}
where $\rho_{\rm s}$ and $r_{\rm s}\equiv r_{\rm vir}/c_{\rm vir}$ are the
scale density and the scale radius, respectively, and $c_{\rm vir}$ is the
concentration parameter. The $r_{\rm vir}$ is calculated from the spherical
collapse model \citep{bryan1998}, and $c_{\rm vir}$ is given by a fitting
formula obtained from N-body simulations (\citealt{Duffy2008}; see also
\citealt{Bullock2001, Wechsler2002, ZhaoD2003, Maccio2007}). The effect of the
evolution of $r_{\rm vir}$ and $c_{\rm vir}$ with redshift will be discussed in
Section~\ref{sec:prob_prob}.  The DM density distribution outside the virial
radius is given by
\begin{eqnarray}
\rho_{\rm DM}(r) & = &  \rho_{\rm DM}(r_{\rm vir}) \left(\frac{r}{r_{\rm
vir}}\right) ^{\delta}\exp\left(-\frac{r-r_{\rm vir}}{r_{\rm decay}}\right),
\nonumber \\ & & \quad \quad \quad \quad \quad {\ \rm if\ } r > r_{\rm vir}.
\label{eq:rhoDMrgt}
\end{eqnarray}
where we follow \citet{Kazantzidis2004} and implement an exponential cutoff
that suppresses the profile on a truncation scale $r_{\rm decay}$ to avoid a
divergent total mass and the parameter $\delta$ is set by keeping the first
derivative of the DM density profiles continuous at $r=r_{\rm vir}$.  The
parameter $\rho_{\rm s}$ is set so that the total mass obtained by integrating
Equations (\ref{eq:rhoDMrltvir}) and (\ref{eq:rhoDMrgt}) over the space is the
total DM mass of the cluster. Given the mass density distribution, the
distribution function of the DM in the six dimensional space $f({\bf r},{\bf
v})$ is assumed to be ergodic and solved via the Eddington's formula (eq.\ 4.46
in \citealt{Binney2008}). In both the SPH and the AMR simulations, DM is
described by Lagrangian particles. To keep the stability of the DM distribution
within the virial radius over cosmological relevant timescales, the truncation
scale is set to $0.3r_{\rm vir}$, as done in \citet{Zemp2008}, and we have
tested that a single cluster with the truncated model is stable within the
Hubble time in our simulations.

We choose the Burkert profile \citep{Burkert1995} to represent the initial gas
density profile as follows,
\begin{equation}
\rho_{\rm gas}(r)=\frac{\rho_{\rm c}}{[1+(r/r_{\rm c})^2](1+r/r_{\rm c})},
{\ \rm if\ } r\le r_{\rm vir},
\end{equation}
as done in \citet{Zuhone2009}, where $r_{\rm c} = 0.5r_{\rm s}$ is the core
radius and the normalization density $\rho_{\rm c}$ is set so that the baryonic mass fraction within $r_{\rm vir}$ is consistent with the cosmological average
value $\Omega_{\rm b}/\Omega_{\rm m}=0.17$. For the region outside the virial
radius, we assume that the gas density profile traces the DM density profile as follows,
\begin{equation}
\rho_{\rm gas}(r)=\rho_{\rm DM}(r)\frac{\rho_{\rm gas}(r_{\rm vir})}{\rho_{\rm
DM}(r_{\rm vir})}, {\ \rm if\ } r> r_{\rm vir}.
\end{equation}

The specific internal energy of the gas at radius $r$ is determined by
\begin{equation}
\mathcal{E}(r)=\frac{1}{\rho_{\rm gas}(r)(\gamma-1)}\int_r^{\infty}\rho_{\rm
gas}(r')\frac{GM(r')}{r'^{2}}dr',
\end{equation}
where the gas is assumed to be in hydrostatic equilibrium and ideal monatomic
gas state with mean molecular weight per ion $\mu=0.592$, $\gamma=5/3$ is the
ratio of the heat capacity at constant pressure to that at constant volume, and
$M(r)$ is the total mass within radius $r$. The temperature distribution of the
gas can be obtained from its internal energy distribution.

We also test different gas density distribution models in our work, e.g., the
$\beta$-model \citep{Cavaliere1978} or a gas distribution tracing the DM
density profile at all radii. We find that the different gas models do not
affect our results significantly, as we focus on the positions of the maxima of
projected X-ray and SZ maps of merging systems.

\subsection{Simulation settings}

In our simulations, the centers of the two clusters with masses $M_1$ and $M_2$
($M_1\ge M_2$) are initially separated by a distance $d_{\rm ini}$ set to twice
the sum of their virial radii, with initial relative velocity $V$ and impact
parameter $P$.  The center of mass of the two clusters is initially put at rest
at the origin of the coordinate system in the simulations. The initial position
of each cluster center is put at the $x-y$ plane with coordinate $z=0$, and
their $(x,y,z)$ coordinates are $(M_2P/(M_1+M_2),\ M_2d_{\rm ini}/(M_1+M_2),\
0)$ and $(-M_1P/(M_1+M_2),\ -M_1d_{\rm ini}/(M_1+M_2),\ 0)$, respectively. The
initial relative velocity is along the $y$-axis, which are $(0,\
-M_2V/(M_1+M_2),\ 0)$ and $(0,\ M_1V/(M_1+M_2),\ 0)$ for the two clusters,
respectively.

We explore a large range of the parameter space of the initial conditions.  We
perform a series of simulations with $M_1=1\times10^{14},\ 2\times10^{14},\
5\times10^{14},\ 1\times10^{15}\msun$. The mass ratio $\xi$($\equiv M_1/M_2$)
is generally set to be 1, 2, 3, or 5. For some special cases, the values of the
mass ratio are set more intensively from 1 to 5. Different from previous work
in which a fixed initial relative velocity (1.1 times circular velocity at the
virial velocity of the larger cluster) is adopted \citep{McCarthy2007,
Zuhone2011}, our simulations are done for a large range of $V$ from $250$ to
$4000 \kms$ and show that the SZ-X-ray offset has a strong dependence on the
relative velocities (see Section \ref{sec:result_map_v} below). The impact
parameter $P$ spans from 0 to $600 \kpc$, as head-on or nearly head-on
mergers are relevant here (see Section \ref{sec:result_map_p} below).

We use the GADGET-2, an efficient parallel TreeSPH code, to carry out our
simulations \citep{Springel2001}. The mass of each gas particle is $m_{\rm
gas}=1.25\times10^8\msun$, and the mass of each dark matter particle is $m_{\rm
DM}=6.25\times10^8\msun$. The gravitational softening length is set to $4 \kpc$, for which other choices (e.g., 1.5 or $15 \kpc$) are also
tested and our main results are not affected.

As mentioned above, we also use the FLASH, a mesh-based Eulerian code, to test
whether the discrepancy between Eulerian and Lagrangian methods affects our
results significantly \citep{Fryxell2000}. The FLASH uses adaptive mesh
refinement (AMR), and solves the equations of hydrodynamics by
Piecewise-Parabolic Method (PPM) of \citet{Colella1984}. The gravitational
potential is computed by the multigrid solver \citep{Ricker2008}. We simulate
the cases with $\xi=2,\ V=500 \kms,\ P=0 \kpc$ for $M_1=2\times10^{14},\
5\times10^{14},\ 1\times10^{15}\msun$. For the box size of $13.0 \Mpc$ of
all the FLASH mergers, the finest resolution achieved in our simulations is
$12.7 \kpc$.

\subsection{Projection analysis: the X-ray and the SZ signals of merging
clusters}

Given any time of the merging processes, we can obtain the two-dimensional maps
of the mass surface density, the X-ray surface brightness, and the SZ emission
of the simulated merging clusters by the following equations.  \begin{itemize}
\item The mass surface density $\Sigma$ is given by an integral of the mass
density along the line of sight (LOS), i.e.,
\begin{equation} \label{eq:density}
\Sigma=\int_{\rm LOS}^{}(\rho_{\rm DM}+\rho_{\rm gas})d\ell.
\end{equation}
\item The X-ray surface brightness is obtained by using an approximate
expression for the relativistic X-ray thermal bremsstrahlung as follows (see
eq.\ 5.25 in \citealt{Rybicki1979}), considering that the gas temperatures
$T_{\rm gas}$ of some regions heated by shocks can be up to $>30\keV$,
\begin{eqnarray}
I_{\rm X} &\propto & \int_{\rm LOS}^{}\rho_{\rm gas}^{2}{T_{\rm
gas}}^{1/2}g_{\rm B} (1+4.4\times10^{-10}T_{\rm gas})d\ell,  \nonumber \\
\label{eq:x-ray}
\end{eqnarray}
where $g_{\rm B}=(2\sqrt{3}/\pi)[1+0.79(4.95\times10^{5}\K/T_{\rm gas})^{1/2}]$
is the frequency average of the velocity averaged Gaunt factor
\citep{Rephaeli1997}.
\item The SZ surface brightness is given by
\begin{eqnarray}
I_{\rm SZ}& \propto & \frac{\sigma_{\rm T}}{m_{\rm e}c^{2}}\int_{\rm
LOS}^{}\rho_{\rm gas} T_{\rm gas} \times \nonumber \\ & &
(Y_{0}+Y_{1}\Theta+Y_{2}\Theta^{2}+Y_{3}\Theta^{3}+Y_{4}\Theta^{4})d\ell,
 \label{eq:sz} \end{eqnarray}
where $\sigma_{\rm T}$ is the Thomson cross section, $m_{\rm e}$ is the
electron mass, $c$ is the speed of light, $\Theta\equiv k_{\rm B}T_{\rm
gas}/(m_{\rm e}c^{2})$, $k_{\rm B}$ is the Boltzmann constant, and $Y_{0},\
Y_{1},\ Y_{2},\ Y_{3},\ Y_{4}$ are the factors given by eqs.~2.26--2.30 in
\citet{Itoh1998}.
\end{itemize}
In practice, we obtain all the above quantities at a given position calculated
from the SPH method by smoothing $N$ neighbor particles output from each
snapshot of the simulations, where $N=50\pm2$ is a neighbor particle number
typically used for smoothing.

The values and the spatial positions at the peaks of the projected mass surface
density, X-ray and SZ images can then be obtained. Note that all the maps have
been smoothed by a Gaussian distribution function with dispersion $\sigma=70 \kpc$.  We also try other values of $\sigma=30,\ 40 \kpc$, and
find that the variations of the positions of the peaks are smaller than $10\%$.
The SZ peak position is not sensitive to the $\sigma$ value even when
$\sigma\geq140 \kpc$. A similar result is reported in
\citet{Molnar2012} although they use a lower SZ resolution.  Thus, not only are
the obtained peak positions not affected by the smoothing, but also they can be
straightforwardly compared with the current SZ observations, though the current
observational resolutions are at the sub-arcminute level.

Finally, we clarify the definition of the ``SZ-X-ray offset'' to be obtained
from our simulations, which is not exactly the same as that obtained from
observations in the following two points. (1) The offset that we obtain from
our simulations is the distance between the positions of the maxima of the
whole X-ray and SZ surface brightness maps, which covers a sufficiently large
region. However, the offsets obtained from observations are sometimes the
displacements between the X-ray and SZ peaks located in a small local region,
hence they can be smaller than the offsets obtained in this work. (2) The
centroids of the galaxy clusters obtained in the X-ray survey are always
obtained by computing the mean emission-weighted positions after removing the
extended secondary X-ray maxima, instead of finding the maximum in the surface
brightness maps as done here \citep[e.g.,][]{Vikhlinin2009, Andersson2011}.

\section{Simulation Results} \label{sec:result}

In this section, we present the following aspects of our simulation results.
(1) We show the projected X-ray and SZ surface brightness maps obtained from
Equations (\ref{eq:x-ray}) and (\ref{eq:sz}), viewed along the $z$-axis unless
otherwise stated. We qualitatively describe the formation and evolution of the
SZ-X-ray offsets during the merging process and discuss the impacts of
different relative velocities, impact parameters, primary masses, and mass
ratios on the maximum and time duration of the offset. (2) We quantitatively
measure the time duration of the SZ-X-ray offset, which is used for estimating
the probability of the offsets expected in observations in Section
\ref{sec:prob}. (3) We also compare the merging processes simulated by the SPH
and the AMR methods.  We demonstrate the robustness of our results obtained by
only using the SPH method.

\subsection{Mapping the X-ray and the SZ emissions from merging clusters}
\label{sec:result_map}

Generally, the merging process of two clusters involves the five distinct
stages \citep{Poole2006}: pre-interaction stage, primary core-core
interaction, apocentric passage, secondary core accretion, and relaxation (see
an example illustrated in Fig.~\ref{pic:offset_v} below), if the initial
relative velocity is not high enough to detach the two clusters from each other
at a later time. As mentioned above, the DM and the gas have different physical
behavior during the mergers. The DM is collisionless so that the DM halos of
the two clusters can go through each other, undergoing only gravity; while gas
experiences gas pressure and shocks can be created during collisions of the gas
halos. The different behavior can be revealed from the images of the mass
surface density, and the X-ray and the SZ emissions of the merging clusters, as
DM dominates the entire mass of the merging system and the X-ray and the SZ
emissions are merely sensitive to the baryon distribution.

\subsubsection{Dependence on initial relative velocities}
\label{sec:result_map_v}

\begin{figure*}%[H]
\centering
   \subfigure[\label{pic:2d_v}]{\includegraphics[scale=0.8]{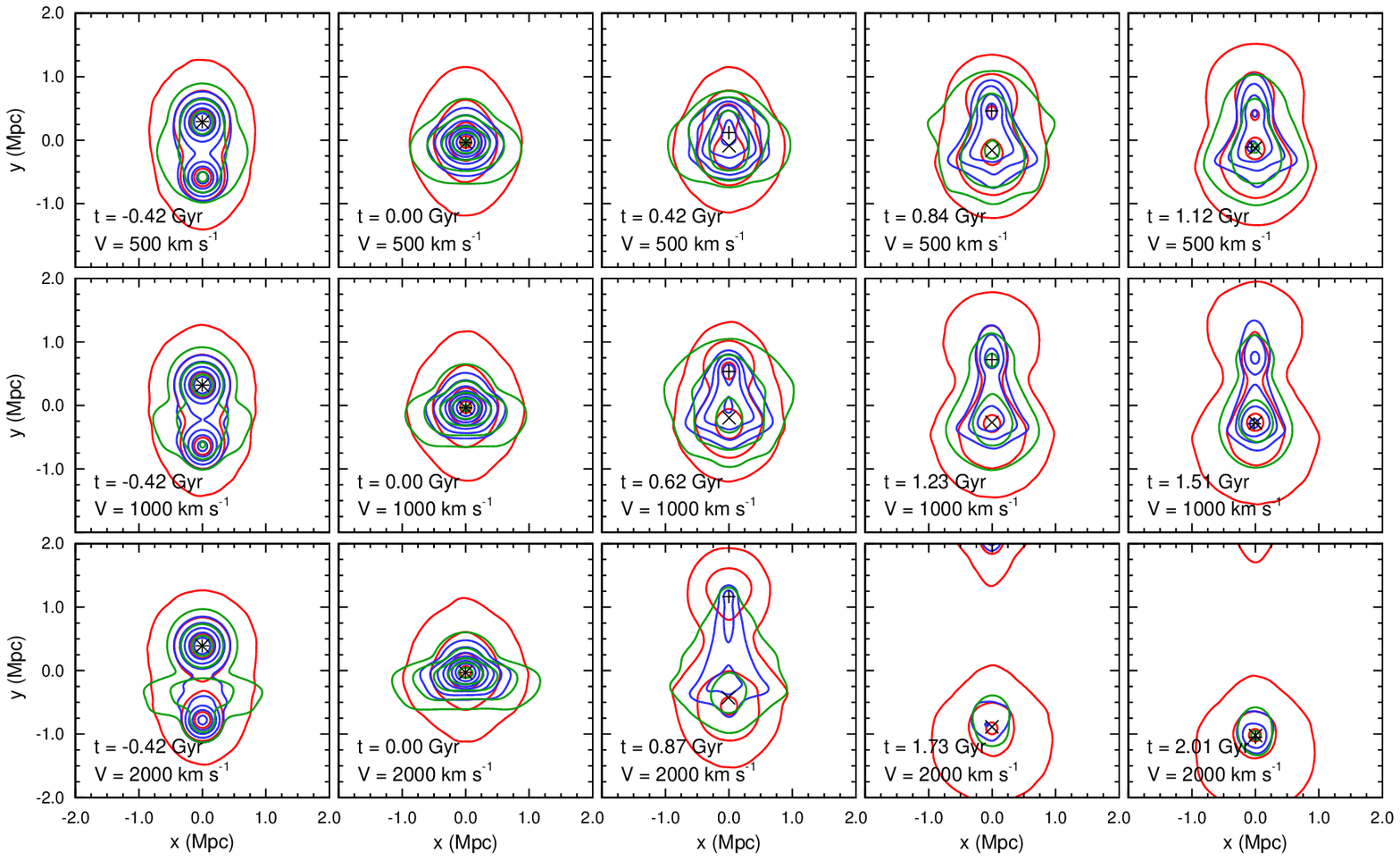}}
   \subfigure[\label{pic:1d_v}]{\includegraphics[scale=0.6]{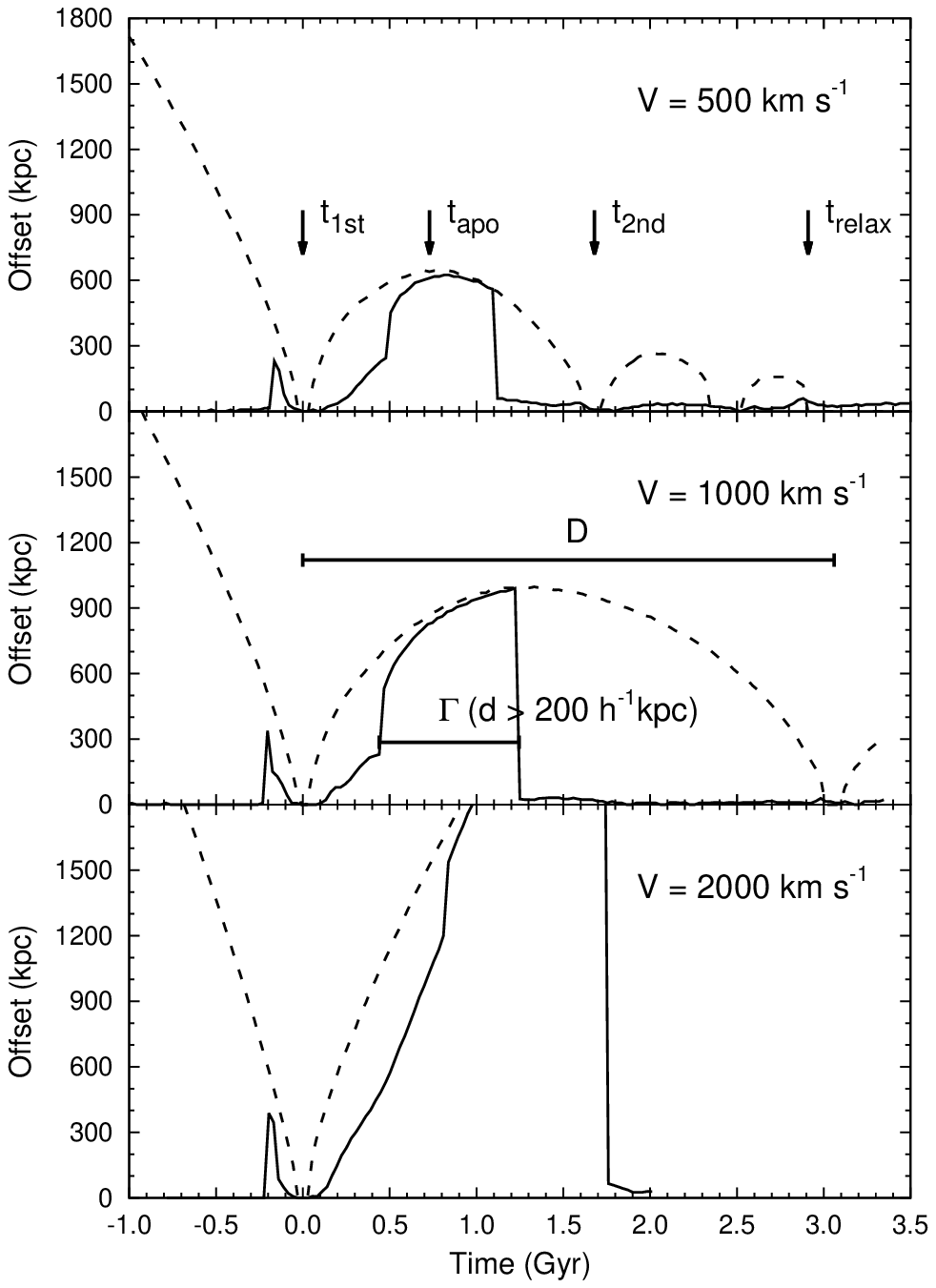}}
\end{figure*}
\begin{figure*}%[H]
\caption{(a) Snapshots of the map of the X-ray and the SZ emissions from two
merging clusters at different merging time $t$. The two clusters have
$M_1=2\times10^{14}\msun$, $\xi=2$, $P=0 \kpc$. The red, green, and blue
contours represent the logarithm of the projected mass surface density, SZ and
X-ray surface brightness, respectively (see
Eqs.~\ref{eq:density}--\ref{eq:sz}); and the ratio in the surface brightness
between successive contour curves is 3.16.  The `$+$' and the `$\times$'
symbols represent the positions of the maxima of the X-ray and SZ maps,
respectively. The second column shows the snapshots at the time of the primary
pericentric passage, where the time is set to $t=0.0\Gyr$. The fourth column
represents the snapshots with the largest SZ-X-ray offset during the merger
processes. (b) Evolution of the SZ-X-ray offsets (solid lines) with the same
initial relative velocities as those in panel (a). The dashed lines are the
distances between the mass density centers of the two clusters as a function of
time. The characteristic time points of different merging stages are marked in
the top panel ($M_1=2\times10^{14}\msun,\ \xi=2,\ P=0 \kpc$): $t_{\rm
1st}$ (primary pericentric passage), $t_{\rm apo}$ (first apocentric passage),
$t_{\rm 2nd}$ (secondary pericentric passage), and $t_{\rm relax}$
(relaxation). The duration of the offset larger than $200 h^{-1}\kpc$ is marked
by $\Gamma(d>d_{\rm c})$ in the middle panel. The symbol ``D'' represents the
time duration between the primary and secondary pericentric passages.  This
figure shows that with higher initial relative velocity, the maximum of the
SZ-X-ray offset appearing after the primary pericentric passages is larger.}
\label{pic:offset_v}
\end{figure*}

Figure~\ref{pic:2d_v} shows a time sequence of the snapshots of the mass
surface density (red contours), the X-ray surface brightness (blue contours),
and the SZ emission (green contours) of a head-on merger of two clusters
($M_{1}=2\times10^{14}\msun$, $\xi=2$) with different initial relative
velocities. For simplicity, we show the results of the first three stages (the
entire five evolution stages can be viewed in Figure~\ref{pic:1d_v}).  The
first column shows a snapshot in the pre-interaction stage for each merging
system.  The second column represents the time of the first pericentric passage
of the two clusters.  For simplicity, we set the evolution time $t=0.0\Gyr$ at
the first pericentric passage.  As seen from the second column, the blue and
the green contours are flattened, i.e., gas is squeezed outwards in the
direction normal to the collision axis; and the green contours are relatively
more flattened, as the outer region of the merging cluster heated by shocks
contributes more to the SZ surface brightness than to the X-ray surface
brightness. The SZ and the X-ray peaks (labeled by `$\times$' and `$+$' in the
figure) start to separate after the first pericentric passage, when the gas at
the inner region of the merging clusters is strongly disturbed. Since the
central mass density of the large cluster is lower than that of the small
cluster, the larger cluster core is penetrated by the small one, and the offset
is gradually stretched (see the third column). The fourth column presents the
snapshots with the largest peak offsets occurred during the merging processes.
The offset returns to a small value or disappears at a later time as the
surrounding gas gradually falls into the gravitational potential of the larger
cluster (see the fifth column).  After the secondary core-core interaction
(though not shown in Figure~\ref{pic:2d_v}; cf., the primary core-core
interaction in the second column), gas is usually partially relaxed. The
SZ-X-ray offsets are then not larger than $100 \kpc$ except for some
special massive systems (e.g., bottom panel in Figure~\ref{pic:1d_m} below).

By comparing the first two rows in Figure~\ref{pic:2d_v}, one can see that
there is no obvious qualitative difference between the evolutionary behavior of
the offset for the case with relative velocity $V=500\kms$ and that with
$V=1000\kms$, except that the largest offset produced during the merging
process is larger for the case with a higher relative velocity.  For the case
with $V=2000 \kms$ (the third row of Figure~\ref{pic:2d_v}), however, the
result is distinctly different as the small cluster escapes away and cannot get
back to collide with the large cluster again after its first pericentric
passage because of the high relative velocity. The center of the small cluster
becomes the maximum of the X-ray brightness map, when it passes through the
larger one. The offset can be up to $3\Mpc$ after their interaction.

\begin{figure}%[H]
\centering
\includegraphics[width=0.4\textwidth]{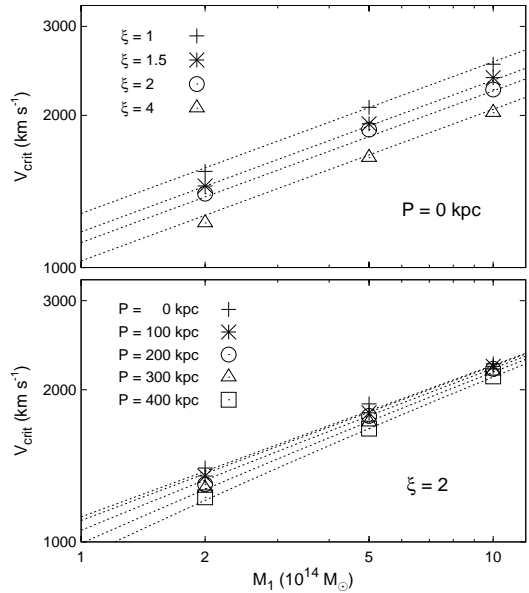}
\caption{Critical velocity $V_{\rm crit}$ as a function of mass $M_1$ for
simulations with different mass ratio $\xi$ (top panel) and impact parameter
$P$ (bottom panel). The points are the simulation results, and the dotted lines
show the fitting results of Equation (\ref{eq:V_c}) for different parameters.
See Section~\ref{sec:result_map_v}.}
\label{pic:V_c}
\end{figure}

Evolution of the SZ-X-ray offsets for different relative velocities is shown in
Figure~\ref{pic:1d_v}. The solid lines give the SZ-X-ray offset as a function
of time. The dashed lines represent the evolution of the distance between the
mass density centers of the two clusters; and a zero distance is used if the
two objects cannot be identified as they overlap or merge together.  The
merging processes are divided into the five different stages as illustrated in
the top panel. Initially, the two clusters are in the pre-interaction stage,
and then the primary core-core interaction starts around the time of the first
pericentric passage at $t_{\rm 1st}(=0)$. Before $t_{\rm 1st}$, it is the
pre-interaction stage.  The time of the first apocentric passage is marked as
$t_{\rm apo}$. After $t_{\rm apo}$, the second core accretion stage starts; and
$t_{\rm 2nd}$ marks the secondary pericentric passage and the end of the second
core accretion stage. As the small cluster in the bound orbit is dragged back
and forth during the gradual relaxation process, the apocentric distances damp
with time.  The whole system appears visually relaxed around the time $t_{\rm
relax}$ as marked in the top panel.

As seen from Figure~\ref{pic:1d_v}, the offset is initially zero.  A small jump
appears before the first pericentric passage, and it then decreases and
disappears as the two clusters get closer. The appearance of the small jump in
the pre-interaction stage can be understood as follows: as the two clusters
approach each other, their outer layers start to be compressed and heated,
during which the position of the maximum SZ signal is affected, but the peak of
the X-ray remains located at the center of the larger cluster.  After the first
pericentric passage, the collision of the two clusters destroys the hydrostatic
equilibrium and spherical symmetry of the gas halos.  As shown in
Figure~\ref{pic:1d_v}, a significantly large SZ-X-ray offset always occurs
between the primary and secondary pericentric passages. The occurrence of the
large offset can be understood as follows: the gas core of the larger cluster
is disrupted, and the X-ray peak jumps to the densest region around the center
of the small cluster. However, the SZ peak does not jump (see the peak
positions shown in the fourth column of Fig.~\ref{pic:2d_v}), as the
temperature at the center of the small cluster is relatively low (see the top
right panel in Fig.~\ref{pic:2d_dens_temp} below), which strongly reduces the
SZ surface brightness whose emissivity is proportional to the temperature. The
SZ-X-ray offset is therefore nearly boosted to the mass density displacement
between the two clusters (hereafter we refer to it as the `jump effect'). As
seen from the figure, the maximum offset is positively correlated with the
initial relative velocity of the two clusters.  The X-ray peaks drop back to
the center of the larger cluster mostly after $t_{\rm apo}$, when the gas falls
back to the trough of the gravitational potential well.  The end of the large
SZ-X-ray offset is indicated by the sharp discontinuity in the solid line. Our
results suggest that the jump effect should be the dominant reason to lead to
the large offset. Our calculations show that adopting different image smoothing
scales $\sigma$ may result in different time durations of the jump effect if
$\sigma>70 \kpc$. The duration of the X-ray peaks locating in the center
of the small cluster is shorter if $\sigma$ is larger, as more substructures in
the X-ray map is smoothed. For example, for the offset larger than $300 \kpc$, the duration obtained with $\sigma=140 \kpc$ is only half of
the value obtained with $\sigma=70 \kpc$.

\begin{figure*}%[H]
 \centering
  \subfigure[\label{pic:2d_p}]{\includegraphics[scale=0.8]{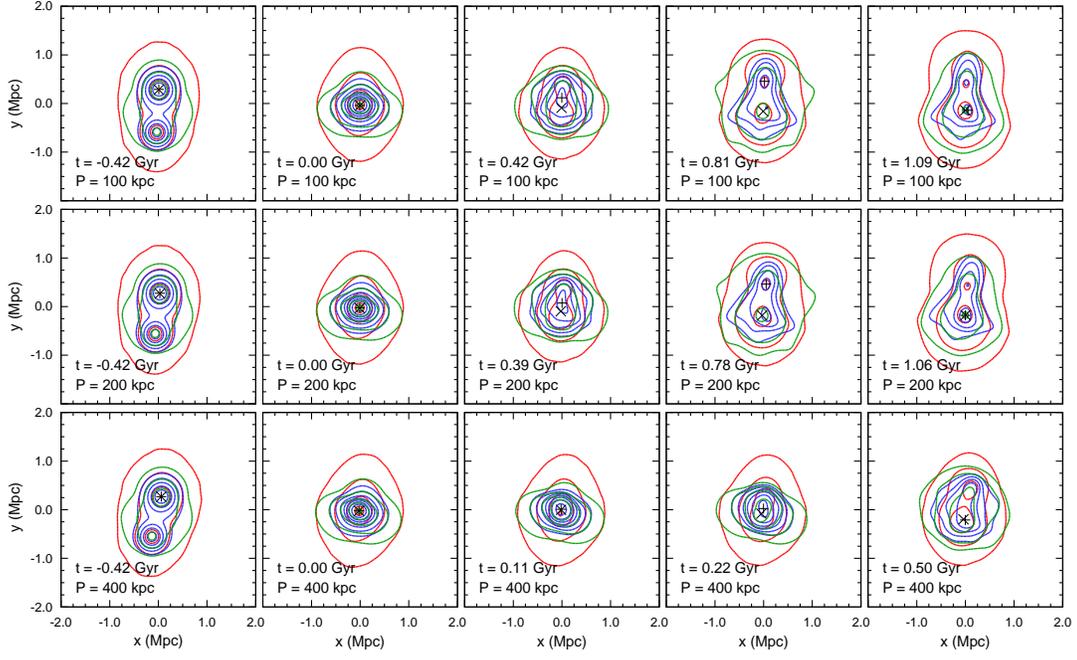}}
  \subfigure[\label{pic:1d_p}]{\includegraphics[scale=0.6]{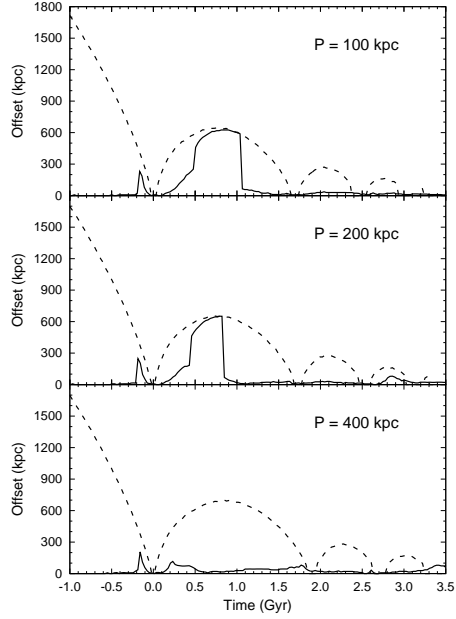}}
\caption{Same as Figure~\ref{pic:offset_v}, but for different impact parameters
with fixed $M_1=2\times10^{14}\msun,\ \xi=2$ and $V=500 \kms$. This figure
shows that the mergers with small impact parameters (top and middle panels) can
produce a large offset (e.g., $>100 \kpc$), while the mergers with large
impact parameters (bottom panels) not. } \label{pic:offset_p}
\end{figure*}

As mentioned above, the evolution of the offset shows different behavior for
low (e.g., $V<1000\kms$) and high (e.g., $V>2000\kms$) relative velocity
mergers.  Here we define a critical velocity $V_{\rm crit}$ to distinguish
these bound and unbound collision cases, i.e., the two clusters will merge and
relax within $13\pm0.5\Gyr$ if $V<V_{\rm crit}$.  The merger is referred to as
the ``merger mode'' if $V<V_{\rm crit}$, otherwise as the ``flyby mode''. We
perform a series of low-resolution simulations (with fixed DM and gas particle
numbers, 40,000 and 10,000 respectively, for computing efficiency) to identify
$V_{\rm crit}$ as a function of $(M_1,P,\xi)$.  The result is shown in
Figure~\ref{pic:V_c}, and the uncertainty in the obtained $V_{\rm crit}$ is
$\pm50\kms$. We approximate the relation between the $V_{\rm crit}$ and the
initial condition parameters by the following fitting form:
\begin{eqnarray} \label{eq:V_c}
  \frac{V_{\rm crit}}{10^3\ \kms}=
A_V\left(\frac{M_1}{10^{14}\msun}\right)^{\alpha_V}(1+\xi^{-1})^{\beta_V}\times
\nonumber \\ \left[1+\gamma_V\left(\frac{P/(100 \kpc)}{(M_1/10^{14}\msun)^{1/3}}\right)^2\right]^{-1/4},
\end{eqnarray}
where the best-fit parameters are $A_V=0.93$, $\alpha_V=0.30$, $\beta_V=0.46$, and $\gamma_V=0.069$.
The critical velocity is strongly related with the primary cluster mass and the
mass ratio, but not the impact parameter.  The fitting form and the best-fit
power-law factors are roughly consistent with the expectation from the escape
velocity criterion in a simple two-body gravitational interacting system, that
is, $V_{\rm crit}^2/2-GM_1(1+\xi^{-1})/d_{\rm ini}=0$, and thus
$V_{\rm crit}\propto M_1^{1/3}(1+\xi^{-1})^{1/2}(1+f_VP^2/M_1^{2/3})^{-1/4}$, where $f_V$ is a factor.

\subsubsection{Dependence on impact parameters} \label{sec:result_map_p}

Figure~\ref{pic:2d_p} shows the snapshots of the simulation results for
different impact parameters with $V=500\kms$.  For the low initial relative
velocities, the large offsets ($>100 \kpc$) occur in both head-on and
nearly head-on impacts (i.e., $P<400 \kpc$). They can reach up to
600--700$\kpc$ (see the top and middle panels in Fig.~\ref{pic:1d_p}).
A larger impact parameter results in a smaller offset, which implies that the
size of the SZ-X-ray displacement is strongly related with the intensity of the
primary core-core interaction. \citet{Molnar2012} studied the high-initial
relative velocity case with $V=4800 \kms$ and find that significant
displacements ($\sim 300 \kpc$) between the SZ and X-ray peaks can be
produced for non-zero impact parameters (about 100--250\kpc) and they decrease
with increasing impact parameters; however, the displacement is insignificant
for zero impact parameter. We also perform simulations with $V=4000 \kms$
similar to those done in \citet{Molnar2012}.  For the non-zero impact parameter
cases, we find that the patterns of the mass surface density, the X-ray, and
the SZ emission are all very similar with those shown in \citet{Molnar2012};
for the head-on merger, however, a displacement up to $150 \kpc$ can
also be produced, while the distance between the mass centers of the two
clusters is $1.5 \Mpc$.

\begin{figure*}
\centering
\includegraphics[scale=0.9]{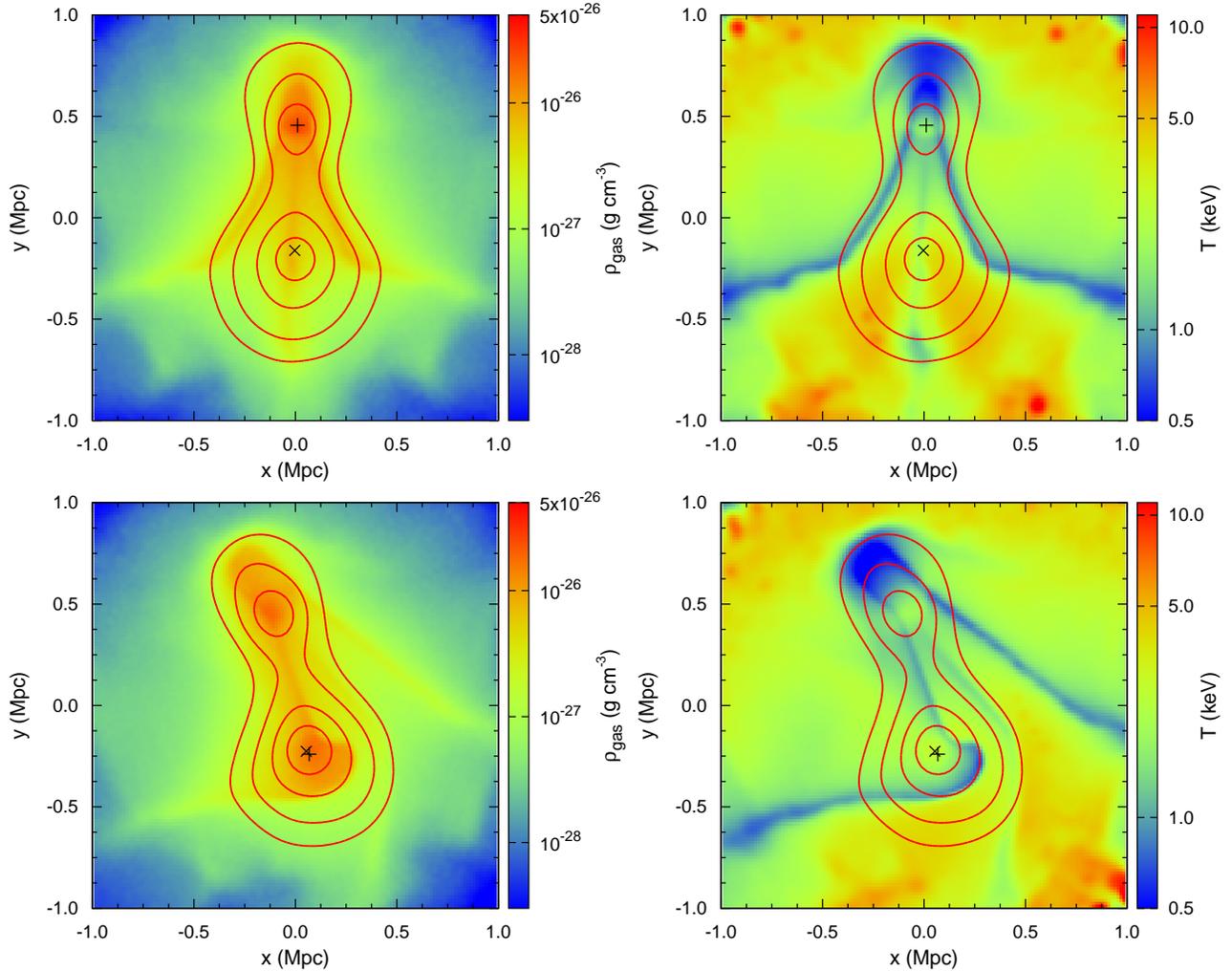}
\caption{Snapshots of the mass density and the temperature slices of the gas at
the coordinate $z=0$ plane at $t=0.75\Gyr$. Top panels are for a head-on
merger with $P=0 \kpc$ and bottom panels for an off-axis merger with
$P=400 \kpc$ ($M_1=2\times10^{14}\msun,\ \xi=2,\ V=500 \kms$). The red
curves are the equi-mass surface density contours of the merging systems. The
`$+$' and the `$\times$' symbols represent the positions of the maxima of the
X-ray and the SZ maps, respectively.  The X-ray peak locates near the center of
the small cluster in the head-on merger, but not in the off-axis one; while the
SZ peaks are close to the center of the large cluster in both cases. The figure
illustrates the reason of the `jump effect' of the SZ-X-ray peak offset for
the head-on merger (see Sections~\ref{sec:result_map_v} and
\ref{sec:result_map_p}).}
\label{pic:2d_dens_temp}
\end{figure*}

Figure~\ref{pic:1d_p} shows the evolution of the SZ-X-ray offsets for different
impact parameters. As seen from the bottom panel of the figure, there is no
significant SZ-X-ray offset for a large impact parameter $P=400 \kpc$.
This can be understood through the density and temperature maps of the
colliding clusters shown in Figure~\ref{pic:2d_dens_temp}.
Figure~\ref{pic:2d_dens_temp} compares the density and temperature slices of
the gas at coordinate $z=0$ plane between the head-on ($P=0 \kpc$) and
the off-axis ($P=400 \kpc$) mergers.  For the head-on merger shown in the
top panels, as discussed above, the core of the large cluster is disrupted at
the collision, and the denser region is near the center of the small cluster;
however, the temperature at the center of the small cluster and its surrounding
region is relatively low. But for the off-axis merger with a large impact
parameter shown in the bottom panels, the gas cores of the two clusters
sideswipe each other at the primary collision, and the center of the large
cluster remains dense; thus the X-ray and SZ peaks are both near the center of
the larger cluster. It is worthy to note that though shocks can heat the gas at
shock fronts to a relatively high temperature (e.g., a few ten keV), the X-ray
and the SZ peaks still locate near the centers of the clusters, as the observed
X-ray and the SZ emission is integrated along the line of sight (see
Eqs.~\ref{eq:x-ray} and \ref{eq:sz}).

Our simulation results demonstrate that only the head-on or nearly head-on
mergers are possible to produce offsets larger than $100 \kpc$, e.g.,
$P<400 \kpc$ for the simulation with $M_1=2\times10^{14}\msun$ or $P<600 \kpc$ for $M_1=5\times10^{14}\msun\ (\xi=2,\ V=500 \kms)$. In addition, a
smaller impact parameter induces a longer time duration of the non-zero offset.

\subsubsection{Dependence on masses} \label{sec:result_map_m}

\begin{figure*}%[H]
\centering
  \subfigure[\label{pic:2d_m}]{\includegraphics[scale=0.8]{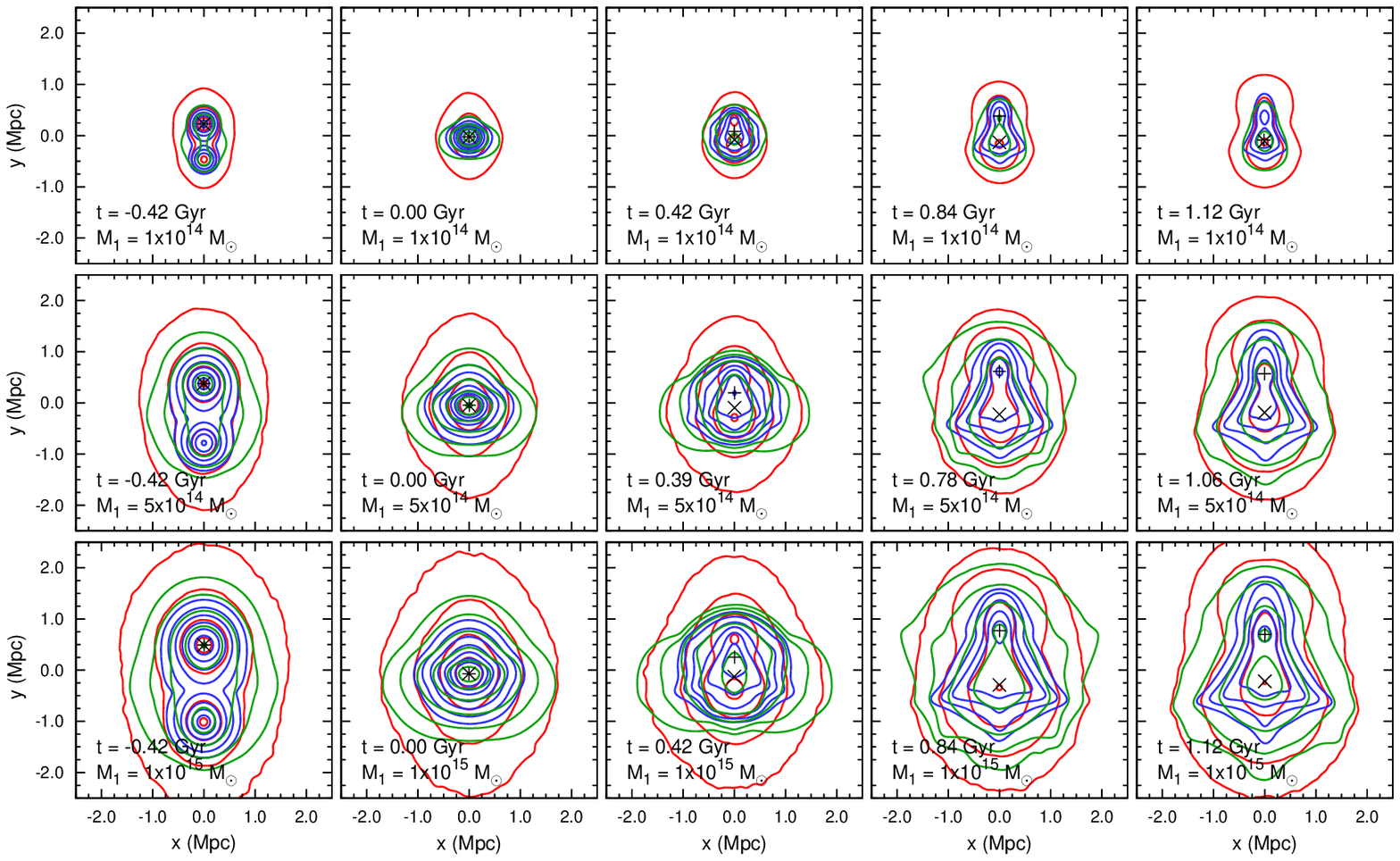}}
  \subfigure[\label{pic:1d_m}]{\includegraphics[scale=0.6]{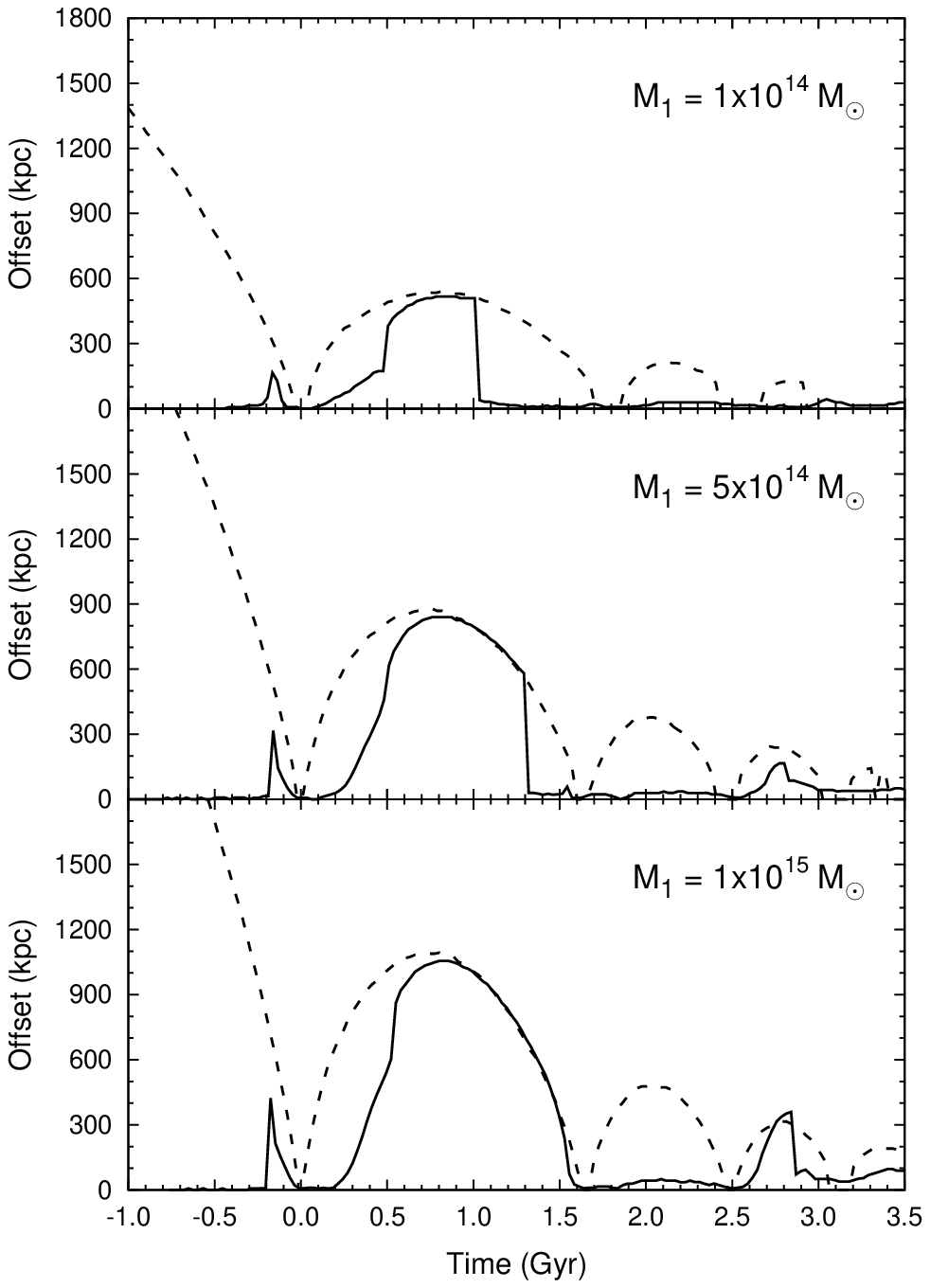}}
\caption{Same as Figure~\ref{pic:offset_v}, but for different masses of the
primary cluster with fixed $\xi=2,\ V=500 \kms$ and $P=0 \kpc$. The
SZ-X-ray offset can be significantly large and  appear in all the cluster mass
range ($>1\times 10^{14}\msun$).} \label{pic:offset_m}
\end{figure*}

Figure~\ref{pic:2d_m} shows the snapshots of the simulations for different
masses of the primary cluster, $M_1=10^{14}$, $5\times10^{14}$, and
$10^{15}\msun$, respectively. Compared to the low-mass case shown in the top
row, the deformation of the gas distribution, especially viewed from the SZ
contours, is much stronger in the more massive merging processes.

Figure~\ref{pic:1d_m} presents the dependence of the SZ-X-ray offset evolution
on $M_1$. As seen from the figure, a significantly large SZ-X-ray offset $(\ga
400 \kpc)$ can occur in the whole mass range of the galaxy clusters. For
the more massive systems, the maximum of the spatial separation between the SZ
and the X-ray peaks is larger, and the time duration of a non-zero offset is
longer.  Our simulation results show that the maxima of the SZ-X-ray offsets
for different masses (denoted by $d_{\rm max}$) are approximately as large as
the first apocentric distances and so does it for different velocities
($V<V_{\rm crit}$). The fitting form to the dependence of $d_{\rm max}$ on the
initial parameters can be obtained first through the following analysis and
then from our numerical simulation results. If approximating dynamics of the
merging system by the dynamics of a two-body system, $d_{\rm max}$ can be
approximately obtained through the energy conservation law
$V^2/2-GM_1(1+\xi^{-1})/d_{\rm ini}\simeq
-GM_1(1+\xi^{-1})/d_{\rm max}$, where $G$ is the gravitational
constant. By setting $\xi=2$ and considering the scaling relation between the
virial radius and the mass of galaxy clusters, we have $d_{\rm ini}=f_d
M_1^{1/3}$ and $d_{\rm max}\simeq d_{\rm ini}(1-V^2d_{\rm
ini}/3GM_1)^{-1} \propto M_1^{1/3}+f_d V^2/3GM_1^{1/3}$, where
$f_d$ is a factor. Thus we fit our numerical simulation results of $d_{\rm
max}$ in the following form,
\begin{eqnarray} \label{eq:d_max}
  d_{\rm max}& = &A_d\left(\frac{M_1}{10^{14}\msun} \right)^{\alpha_d}+
\nonumber \\ & & B_d\left(\frac{V}{10^3 \kms}\right)^2\left(\frac{M_1}{10^{14}
\msun}\right)^{-1/3},
\end{eqnarray}
and obtain the best-fit parameters $A_d=366 \kpc$, $B_d=653 \kpc$,
$\alpha_d=0.42$. Equation (\ref{eq:d_max}) is used in the integration of
Equation (\ref{eq:N_offset}) to obtain the offset rate. Here we do not consider
the dependence of $d_{\rm max}$ on impact parameters because (1) $d_{\rm max}$
is not sensitive to the impact parameter (e.g., $P<400 \kpc$ in Fig.
\ref{pic:1d_p}) unless $P$ is too large to suppress the large offset, and (2)
the cases with large impact parameters do not contribute much to the
integration of Equation (\ref{eq:N_offset}) below, as their time durations of
the non-zero offsets are shorter (or the $\langle S_{\rm p}\rangle$ term in
Eq.~\ref{eq:N_offset} is zero when $P$ is too large).

As seen from the middle and the bottom panels of Figure~\ref{pic:1d_m}, for the
mergers of high-mass clusters (e.g., $M_{1}=5\times10^{14}$, $1\times10^{15}
\msun$), a significantly large offset can still appear after the third
pericentric passage (where the SZ peak deviates from the center of the large
cluster due to an offset between the positions of the maxima of the gas density
and the temperature distributions). However, when doing the statistical
analysis in Section \ref{sec:prob} below, we consider the offsets ($>50
h^{-1}\kpc$) triggered merely between the primary and secondary pericentric
passages for the following reasons: (1) massive mergers with cluster masses
larger than $5\times10^{14}\msun$ are rare events in the universe, which
approximately occupy $5\%$ among all major mergers of galaxy clusters; and (2)
the time duration of the offset after the third pericentric passage is nearly
five times smaller than that of the offset triggered by the first core-core
collision.

\subsubsection{Dependence on mass ratios} \label{sec:result_map_mr}

The mass ratio $\xi$ is also an important parameter to affect the values of the
offsets.  According to our simulations, if the initial relative velocity is
below $2000\kms$ for $M_1=2\times10^{14}\msun$, we find that only when
$1<\xi<3$ (i.e., major mergers) can the offsets be larger than $100 \kpc$. If the mass ratio is larger, the strength of the collision is
weaker and has less power to disturb the large cluster, and thus the offset
becomes less significant. For producing the offset larger than $100 \kpc$
with higher $\xi$, more massive merging systems (e.g., $M_1>5\times10^{14}
\msun$) or higher relative velocities are required. Note that when the mass
ratio approaches unity, the offset also turns to be insignificant, as the
merging configuration is symmetric.

The dependence of the SZ-X-ray offset on the mass ratio provides a
complementary method to constrain the kinematics of an individual merging
cluster (see Section \ref{sec:conclusion} below).  For example, the Bullet
Cluster is a system with two merging clusters with quite different masses; and
by doing the simulations with $M_1=5\times10^{14}\msun,\ \xi=8,\ P=0$, we find
that only when $V>3000 \kms$ can the maximum of the offset be larger than $150 \kpc$.

\subsection{Duration of the SZ-X-ray offset and the offset ratio}
\label{sec:result_ratio}

The time duration of the SZ-X-ray offset is important in this work, as it is
directly associated with the probability of the offset appearing in the
observation. As studied in Section \ref{sec:result_map}, the duration has a
strong relation with the initial parameters $V,\ P,\ M_{1}$ and $\xi$. To
quantitatively describe the duration, we define the ``offset ratio''
$R(d>d_{\rm c})$ by
\begin{equation} \label{eq:offset_ratio}
  R(d>d_{\rm c})\equiv\frac{\Gamma(d>d_{\rm c})}{D},
\end{equation}
where $D\equiv t_{\rm 2nd}-t_{\rm 1st}$ is the time duration between the
primary and the secondary pericentric passages. $\Gamma(d>d_{\rm c})$ is the
time duration of the SZ-X-ray offset larger than $d_{\rm c}$ (e.g., see
$\Gamma$ and $D$ marked in the middle panel of Figure~\ref{pic:1d_v} where
$d_{\rm c}=200 h^{-1}\kpc$)\footnote{Hereafter we use the unit of $h^{-1}\kpc$
for $d_{\rm c}$ for comparison with the observation (see Section
\ref{sec:prob}).}.

The time duration $D$ obtained from our simulations can be fit as
a function of the pairwise velocity, cluster mass, and redshift by the
following form,
\begin{eqnarray} \label{eq:D}
D& = &A_D\left[1-B_D\left(\frac{V/10^3\kms}{(M_1/10^{14}\msun)^{1/3}}\right)^2\right]^{-3/2}
\nonumber \\ & & \times (1+z)^{\alpha_D},
\end{eqnarray}
where $A_D=1.53\Gyr$, $B_D=0.62$, $\alpha_D=-0.77$. The relaxation timescale of
the merging clusters in our simulations (i.e.,
$t_{\rm relax}-t_{\rm 1st}$) is about several Gyr, which is typically $2-3$
times longer than $D$ (e.g., see Figs~\ref{pic:1d_v}, \ref{pic:1d_p}, and
\ref{pic:1d_m}). The time duration of the mergers listed in
\citet{Poole2006} also gives a similar result. In this work, we assume $t_{\rm
relax}-t_{\rm 1st}=\kappa D$ with the factor $\kappa=2$. We will present our
detailed studies of the relaxation timescale in a separate paper.

\begin{figure*}%[H]
  \centering \includegraphics[scale=0.9]{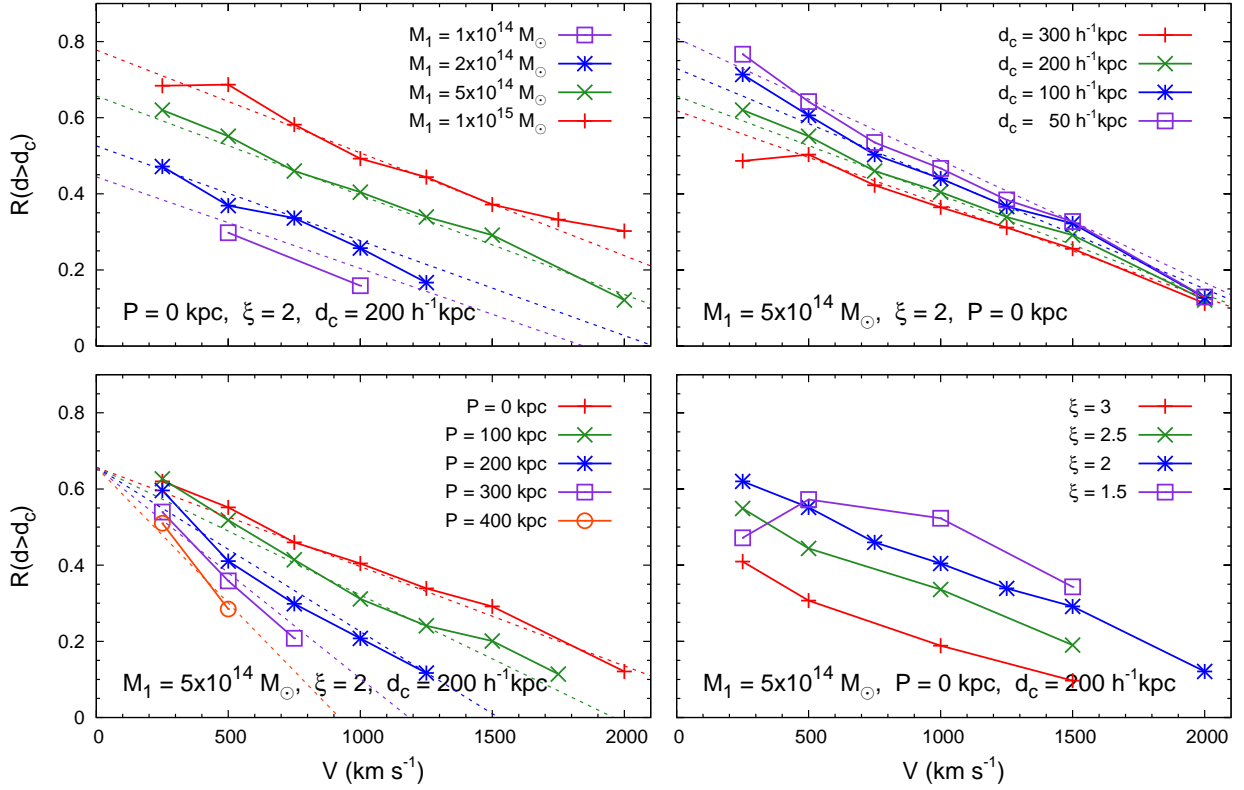}
\caption{Offset ratios $R(d>d_{\rm c})$ as a function of the initial relative
velocity for different initial conditions (see the points with different colors
and the connecting solid lines).  The dotted lines represent the corresponding
best-fit results from Equation (\ref{eq:fit_R1}).}
\label{pic:offset_ratio}
\end{figure*}

In Figure~\ref{pic:offset_ratio}, we present the offset ratio $R(d>d_{\rm c})$
measured from mergers with different initial conditions. We select $d_{\rm
c}=50,\ 100,\ 150,\ 200,\ 300 h^{-1}\kpc$, which fall into the typical offset
range in observations. We only consider the ``merger mode'' whose $D$ is
shorter than $6 \Gyr$ in our estimation for its dominant contribution to the
large offset in the observation (see Section \ref{sec:prob_rate_v} below).
Except for the massive mergers (see Section \ref{sec:result_map_m}), the
offsets larger than $50 h^{-1}\kpc$ appear only between the primary and
secondary collisions, consequently we have $0\le R(d>d_{\rm c})<1$.

The top left panel of Figure~\ref{pic:offset_ratio} shows the offset ratio as a
function of the relative velocity with different $M_1$ for $d_{\rm c}=200
h^{-1}\kpc$, which indicates that mergers of more massive clusters produce
higher offset ratios. For each mass, the ratios reveal an anti-correlation with
the relative velocity, and the correlation slopes for different masses are
close.  The top right panel presents the dependence of the ratio on $d_{\rm c}$
for $M_1=5\times10^{14} \msun$, where the duration of the larger offset is
shorter. However, the difference is not significant at the high-velocity end.
The bottom left panel presents the dependence on different impact parameters
for $M_1=5\times10^{14} \msun$. As seen from the panel, the slopes of the
curves decrease as the impact parameters increase, and consequently the
effective velocity range for a positive offset ratio decreases strongly, e.g.,
$V<1000 \kms$ for $P=400 \kpc$.  The bottom right panel shows the impact
of the different mass ratios on the offset ratio for $M_{1}=5\times10^{14}
\msun$ and $P=0 \kpc$. We find that if $1.5<\xi<2.5$, the amplitudes of
the offset ratios are close; while if $\xi$ is larger than 3 or approaches to
1, the offset ratio decreases significantly.

According to the dependence behavior of the offset ratio on the initial
conditions shown in Figure~\ref{pic:offset_ratio}, we find that the following
functional form fits the data well,
\begin{eqnarray}
    R(d>d_{\rm c}) &= & A_{\rm l}\cdot D_{\rm scale}^{\alpha_{\rm l}}\cdot
\left[\left(\frac{V}{10^3 \kms}\right) \right. \cdot \nonumber \\ & &  \left.
\exp\left(\delta_{\rm l}\cdot P_{\rm scale}\right) - \beta_{\rm
l}\left(\frac{M_{1}}{10^{14} \msun}\right)^{\gamma_{\rm l}}\right], \nonumber
\\
 & &  {\rm for} \ (d_{\rm c}\geq50 h^{-1}\kpc), {\ \rm and} \nonumber \\
& & D_{\rm scale}  =  \frac{d_{\rm c}/100 h^{-1}\kpc}{(M_{1}/10^{14}
\msun)^{1/3}}, \nonumber \\ & & P_{\rm scale}  =  \frac{P/100
\kpc}{(M_{1}/10^{14} \msun)^{1/3}},
\label{eq:fit_R1}
\end{eqnarray}
where the offset ratio is linearly correlated with the initial relative
velocity. The mass-scaled terms $D_{\rm scale}$ and $P_{\rm scale}$ are
included to indicate the dependence on $d_c$ and the impact parameter. We fit
simultaneously to all of the simulation results but fix $\xi=2$. Considering
the limited simulation test explored in this study, the mass ratio is not taken
as an argument in the above fitting formula. The possible effects caused by
different $\xi$ will be discussed at the end of Section~\ref{sec:prob_prob}
below. The best-fit parameters ($A_{\rm l},\alpha_{\rm l},\beta_{\rm
l},\gamma_{\rm l},\delta_{\rm l}$) are given in Table \ref{tab:fit_r}. Note
that Equation (\ref{eq:fit_R1}) is merely suitable for $M_1>10^{14} \msun$ and
$d_{\rm c}=50-300 h^{-1}\kpc$. And the ratio $R(d>d_{\rm c})$ is constrained to
be non-negative. Therefore, the effective velocity range for the positive
offset ratio is $V<\beta_{\rm l}(M_1/ 10^{14}\msun)^{\gamma_{\rm l}}\times10^3
\kms$ for $P=0$, and the upper limit of the effective velocity range is a few
tens percents larger than the critical velocity $V_{\rm crit}$ given in
Equation (\ref{eq:V_c}).  However, when $P>0$, the effective velocity could be
much smaller than $V_{\rm crit}$.

We investigate those cases with offsets smaller than $50 h^{-1}\kpc$ below,
separately, which is necessary especially when we estimate the observational
expectation of the SZ-X-ray offset distribution over all ranges of the offset
size in Section \ref{sec:observation}.  There are at least the following
several reasons to separately investigate the large offsets (i.e., $d_{\rm
c}\geq50 h^{-1}\kpc$) and the small ones (i.e., $0<d_{\rm c}<50 h^{-1}\kpc$),
respectively. (1) As discussed above, the two ranges of the offsets are caused
by different reasons. The large offsets are strongly affected by the `jump
effect', mostly appearing between the primary and secondary pericentric
passages. On the contrary, the small offsets can be viewed during the whole
merging processes. (2) As the uncertainty in our estimation is about a few kpc,
the time duration of the offsets smaller than $50 h^{-1}\kpc$ has a relatively
larger error. (3) Furthermore, the statistics of the offsets smaller than $50
h^{-1}\kpc$ performs relatively irregular behavior, and is more complex to be
described.  Note that the criterion to separate the large and the small offsets
is at $50 h^{-1}\kpc$ or a few dozens of kpc, which is reasonable because the
core radius of the initial gas distribution is $\sim 100 h^{-1}\kpc$.  While
the size of the SZ-X-ray offset is comparable with or larger than the core
radius, the SZ and the X-ray peaks actually locate near the centers of the big
and the small clusters, respectively. Only the disturbed core region of the
big cluster itself could not generate such a large offset.

We count the ratio of the small offsets by using our simulations and setting
$d_{\rm c}=10,\ 20,\ 30,\ 40 h^{-1}\kpc$, respectively.  We find that the
offset ratios also show a linear correlation with the relative velocity as
those of $d_{\rm c}\geq50 h^{-1}\kpc$, though the correlation has a larger
scatter than that shown in Figure~\ref{pic:offset_ratio}. In addition, for
$d_{\rm c}<50 h^{-1}\kpc$, the offset ratios are not strongly related with the
impact parameter, and thus we use a mild dependence on the impact parameter in
the fitting function of $R(d>d_{\rm c})$ given by
\begin{eqnarray}
    & & R(d>d_{\rm c}) =  A_{\rm s}\cdot D_{\rm scale}^{\alpha_{\rm s}}\cdot
\nonumber \\
   & &  \left[\left(\frac{V}{10^3 \kms}\right)-\beta_{\rm s}\cdot D_{\rm
scale}^{\gamma_{\rm s}} \cdot \exp(\delta_{\rm s}\cdot D_{\rm scale}\cdot
P_{\rm scale})\right], \nonumber \\
  & & \quad \quad \quad \quad \quad \quad \quad \quad \quad {\rm for\ } d_{\rm
c}<50 h^{-1}\kpc,
  \label{eq:fit_R2}
\end{eqnarray}
where the best-fit parameters ($A_{\rm s},\alpha_{\rm s},\beta_{\rm
s},\gamma_{\rm s},\delta_{\rm s}$) are given in Table \ref{tab:fit_r}.
\begin{table}%[H]
\begin{center}
\caption{Best-fit parameters for Equations (\ref{eq:fit_R1}) and
(\ref{eq:fit_R2}).}
\label{tab:fit_r}
\begin{tabular}{cccccc}
  \hline \hline
Eq.~(\ref{eq:fit_R1}) & $A_{\rm l}$ & $\alpha_{\rm l}$ & $\beta_{\rm l}$ &
$\gamma_{\rm l}$ & $\delta_{\rm l}$ \\
 &  $-0.267$ & -0.150 & 1.847 & 0.193 & 0.430 \\
  \hline
Eq.~(\ref{eq:fit_R2}) & $A_{\rm s}$ & $\alpha_{\rm s}$ & $\beta_{\rm s}$ &
$\gamma_{\rm s}$ & $\delta_{\rm s}$ \\
 &  $-2.10$ & 0.756 & 0.390 & -1.04 & -0.672 \\
  \hline \hline
\end{tabular} \end{center}
\end{table}

\subsection{Comparison between the SPH and AMR simulations}
\label{sec:result_sphamr}

\begin{figure*}%[H]
\centering
  \subfigure[\label{pic:2d_sphamr_dens}]{\includegraphics[scale=0.8]{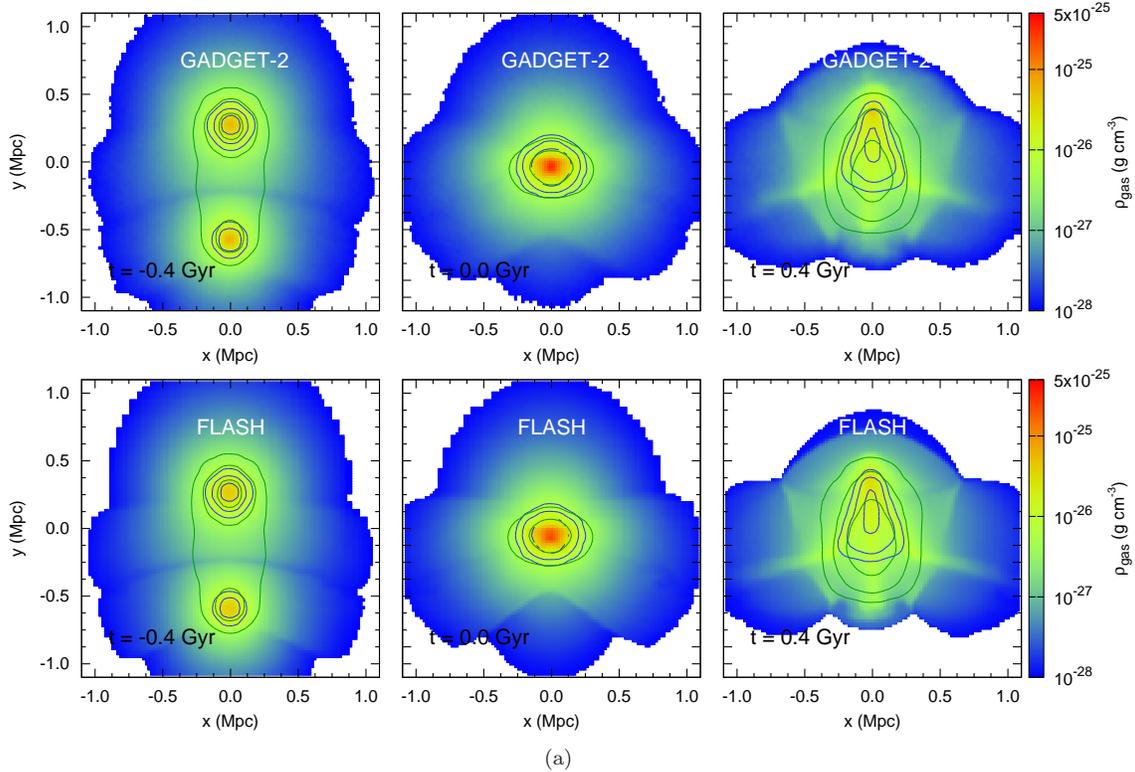}}
  \subfigure[\label{pic:2d_sphamr_temp}]{\includegraphics[scale=0.8]{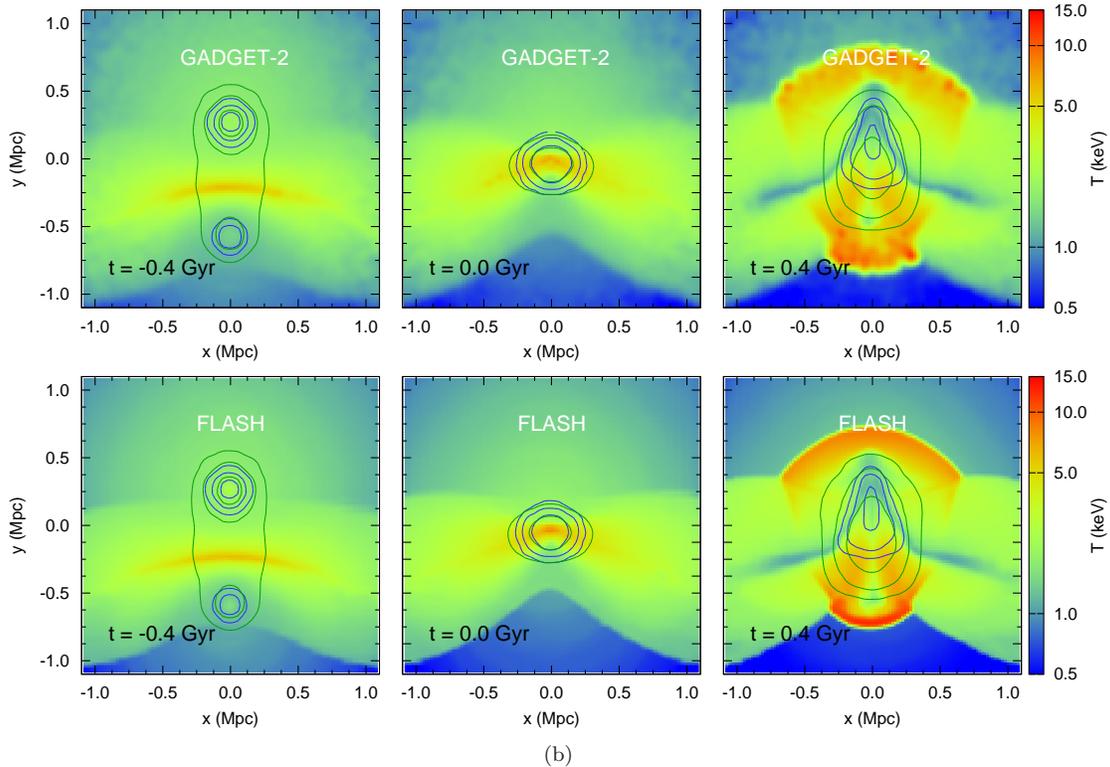}}
%
%\end{figure*}
%
%\begin{figure*}%[H]
%
\caption{The gas mass density distribution (panel a) and the temperature
distribution (panel b) of two merging clusters at the merger plane at different
merging time $t=-0.4,\ 0.0,\ 0.4 \Gyr$. The values of the density and the
temperature distributions are represented by the color scales. The two merging
clusters have $M_1=2\times10^{14} \msun$ and $\xi=2$, with initial relative
velocity $V=500 \kms$ and zero impact parameter. In each panel, the top row
shows the results of GADGET-2, and the bottom row is for those of FLASH. The
overlaid blue and the green contours represent the X-ray and the SZ surface
brightness viewed along the $z$-axis, respectively. For clarity, different
contour levels are applied at different time. This figure shows that the
positions of the X-ray and SZ peaks are affected little by using different
simulation methods, which supports the use of the SPH code in a large number of
the simulations in this work.} \label{pic:2d_sphamr}
\end{figure*}

In the last part of Section \ref{sec:result}, we show an example of simulating
the merging process of two clusters by using both the SPH and the AMR methods.
The parameter settings and the results of the simulation are shown in
Figure~\ref{pic:2d_sphamr}. Figure~\ref{pic:2d_sphamr} shows the slices of the
simulated gas mass density and temperature distributions at the merger plane,
i.e., $\rho_{\rm gas}(x,y,z=0)$ and $T(x,y,z=0)$, obtained at different merging
time. The top and the bottom panels represent the results obtained by the
GADGET-2 and FLASH codes, respectively. The overlaid contours show the
projected X-ray (blue) and SZ (green) surface brightness maps.  As seen from
the figure, the density and the temperature distributions of the merging
structure obtained from the two codes are consistent in general, except for the
two main different points below. (1) The discontinuity produced by shocks are
sharper in the FLASH merger, which reveals the advantage of the AMR method in
capturing sharp gas features. (2) In the inner region of the system, the
density of the gas core produced by the SPH code tends to be higher (by
$<5\%$). This minor discrepancy is suggested to be due to the suppression of
the turbulence by the SPH method, as reported in \citet{Mitchell2009}. These
deviations in the strength of the surface brightness, however, affect the
positions of the X-ray and SZ peaks little, which guarantee the robustness of
our work results obtained by only using the SPH code.

\section{Probability of the SZ-X-ray Offset} \label{sec:prob}

By exploring the parameter space of the initial conditions in Section
\ref{sec:result}, we find that the cluster mergers with $M_{1}>10^{14} \msun,\
P<400 \kpc$ and $\xi\leq3$ may form the SZ-X-ray offset larger than $100 \kpc$. The parameter space almost covers the whole mass range of the
galaxy clusters which experienced major mergers.  In this section, we
investigate the probability of a significant SZ-X-ray offset occurring in
observations.

\subsection{Model of the offset rate} \label{sec:prob_rate}

The number of the mergers with observational offset $d_{\rm phy}>d_{\rm c}$
\footnote{To distinguish from the SZ-X-ray offset $d$ viewed along the
$z$-axis, we use $d_{\rm phy}$ to represent the offset viewed along from any
one randomly given direction.} per unit redshift (or per unit cosmic time after
multiplying by $dz/dt$) per unit comoving volume at redshift $z$, which is
referred to as the offset rate hereafter, over a given range of mass
$M_0\in[M_{\rm min},\ M_{\rm max}]$ ($M_0\equiv M_1(1+\xi^{-1})$, is the total
mass of the merging system), mass ratio $\xi\in[\xi_{\rm min},\ \xi_{\rm
max}]$, impact parameter $P\in[P_{\rm min},\ P_{\rm max}]$, and initial
relative velocity $V\in[V_{\rm min},\ V_{\rm max}]$ can be given by:
\begin{eqnarray}
    \frac{dN(d_{\rm phy} >  d_{\rm c}|z)}{dz}  & = & \int_{M_{\rm min}}^{M_{\rm
max}}dM_0 \int_{\xi_{\rm min}}^{\xi_{\rm max}}d\xi \int_{P_{\rm min}}^{P_{\rm
max}} dP  \nonumber \\
& &  \cdot f_P(P) \int_{V_{\rm min}}^{V_{\rm max}} f_V(V) \cdot H(d_{\rm
max}-d_{\rm c}) \nonumber \\
& &\cdot \langle S_{\rm p}(d_{\rm phy}>d_{\rm c}|M_0,\ \xi,\ P,\ V)\rangle
\nonumber \\
& &  \cdot B(M_0,\ \xi,\ z-\Delta z)\ dV,
 \label{eq:N_offset}
\end{eqnarray}
where $f_P(P)$ and $f_V(V)$ are the initial distribution functions of the
impact parameters and the relative velocities of merging cluster pairs with
$\int f_P(P)dP=1$ and $\int f_V(V)dV=1$, respectively. In Equation
(\ref{eq:N_offset}), $S_{\rm p}(d_{\rm phy}>d_{\rm c}|M_0,\ \xi,\ P,\ V)$
is defined as the specific probability of a merging system with a given initial
condition ($M_1,\ \xi,\ P,\ V$) showing the observed offset $d_{\rm phy}>
d_{\rm c}$ in one observational direction, which can be obtained through the
ratio of the time duration of the observed offset ($>d_{\rm c}$) over the total
merging time (from the primary pericentric passage to the complete relaxation;
cf., Eq.~\ref{eq:specificP} below), and $\langle S_{\rm p}(d_{\rm phy}>d_{\rm
c}|M_0,\ \xi,\ P,\ V)\rangle$ is the average specific probability of the
observed SZ-X-ray offset $d_{\rm phy}>d_{\rm c}$ over all random observational
directions. The $H(d_{\rm max}-d_{\rm c})$ is the Heaviside step function,
where $d_{\rm max}$ is the maximum of the SZ-X-ray offset during the merger
(i.e., Eq.~\ref{eq:d_max}).  The merger rate $B(M_0,\ \xi,\ z)$ is defined so
that $B(M_0,\ \xi,\ z)dM_0 d\xi dz$ represents the comoving number density of
galaxy cluster mergers completed at redshift $z\rightarrow z+dz$, with primary
mass in the range $M_0\rightarrow M_0+dM_0$ and mass ratio in the range
$\xi\rightarrow \xi+d\xi$.  The $\Delta z$ is obtained by the cosmic time
difference $t(z-\Delta z)-t(z)= (t_{\rm relax}-t_{\rm 1st})-\frac{1}{2}(t_{\rm
2nd}-t_{\rm 1st}) $ for large offsets (e.g., $\ga 50h^{-1} \kpc$) and
$t(z-\Delta z)-t(z)=\frac{1}{2}(t_{\rm relax}-t_{\rm 1st})$ for small offsets
(e.g., $\la 50h^{-1} \kpc$), where the large offsets occur mainly at the
primary core-core interaction stage as mentioned in
Section~\ref{sec:result_map_m}.

Below we present the detailed forms of the functions $\langle S_{\rm
p}\rangle$, $B$, $f_V$, and $f_P$ in
Sections~\ref{sec:prob_rate_specificP}--\ref{sec:prob_rate_p}, respectively.

\subsubsection{The average specific probability of the offset $\langle S_{\rm
p}\rangle$} \label{sec:prob_rate_specificP}

\begin{figure}%[H]
  \centering
  \includegraphics[scale=0.8]{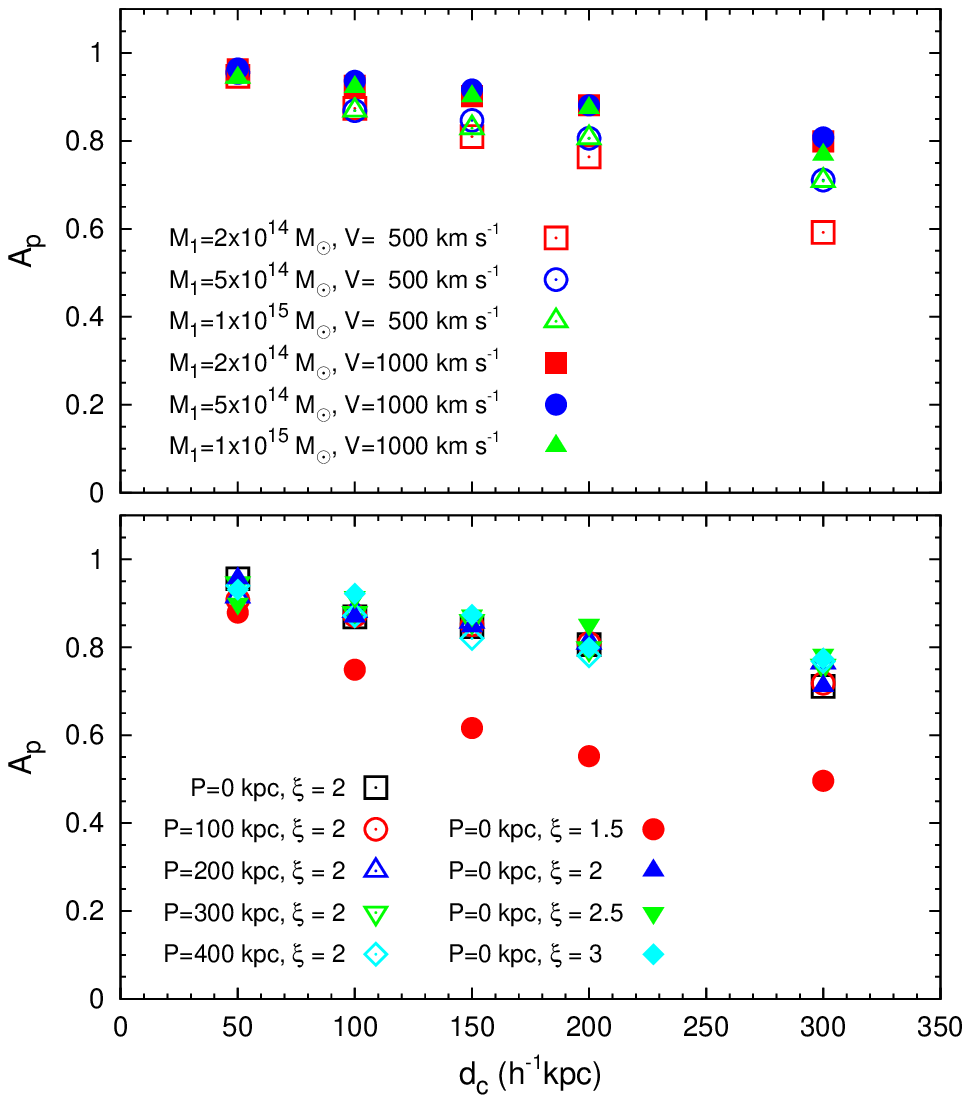}
\caption{The projection factor defined by Equation (\ref{eq:project_factor}) as
a function of $d_{\rm c}$. Different point colors and types represent different
initial parameters. $P$ and $\xi$ are fixed to be $0 \kpc$ and 2 in the
top panel; $M_1$ and $V$ are fixed to be $5\times10^{14} \msun$ and $500 \kms$,
respectively, in the bottom panel. As seen from the figure, the factor does not
significantly depend on the initial conditions.}
\label{pic:projection}
\end{figure}

We define the average specific probability of offset in Equation
(\ref{eq:N_offset}) by
\begin{equation} \label{eq:specificP}
  \langle S_{\rm p}(d_{\rm phy}>d_{\rm c}|M_0,\ \xi,\ P,\ V)\rangle\equiv
\frac{\langle\Gamma(d_{\rm phy}>d_{\rm c})\rangle}{t_{\rm relax}-t_{\rm 1st}},
\end{equation}
where $\langle\Gamma(d_{\rm phy}>d_{\rm c})\rangle$ is the average of the time
duration of the observed offsets larger than $d_{\rm c}$ over all possible
observational directions.  To connect the average duration and the duration
observed along the $z$-axis that was discussed in Section \ref{sec:result}, we
introduce the following projection factor
\begin{equation} \label{eq:project_factor}
  A_{\rm p}(d_{\rm phy}>d_{\rm c}|d>d_{\rm c},\ M_{1},\ \xi,\ P,\
V)\equiv\frac{\langle\Gamma(d_{\rm phy}>d_{\rm c})\rangle}{\Gamma(d>d_{\rm
c})}.
\end{equation}
In principle, the projection factor should depend on the initial parameters and
$d_{\rm c}$. We investigate the dependence by using the Monte-Carlo method and
randomly selecting the observational directions for the given snapshots, and
the result is shown in Figure~\ref{pic:projection}.

As seen from Figure~\ref{pic:projection}, the values of the points at a given
$d_{\rm c}$ are quite close. The only outliers are the red points in the
lower panel, which are for the case with $M_{1}=5\times10^{14}\msun$,
$\xi=1.5$, $V=500\kms$ and $P=0 \kpc$.  The factors for the mergers with
$V=500 \kms$ is around $10\%$ smaller than that for the corresponding cases
with $V=1000 \kms$. That is, the projection factor does not significantly
depend on the cluster mass, the mass ratio, the impact parameter, and the
initial relative velocity. Thus, we assume a universal factor $\langle A_{\rm
p}(d_{\rm c})\rangle$, which is only a function of $d_{\rm c}$. For simplicity,
we set $\langle A_{\rm p}(d_{\rm c})\rangle$ to be the mean value of the data
shown in the top panel of Figure~\ref{pic:projection} at each $d_{\rm c}$.  For
the offsets smaller than $50 h^{-1}\kpc$, we also find a similar result that
the projection factor can be assumed to be only a function of $d_{\rm c}$.  The
results of $\langle A_{\rm p}(d_{\rm c})\rangle$ are listed in Table
\ref{tab:projection}.

\begin{table} %[H]
\begin{center}
\caption{The projection factor $\langle A_{\rm p}(d_{\rm c})\rangle $
(Section~\ref{sec:prob_rate_specificP}). }
 \label{tab:projection}
\begin{tabular}{cccccc}
  \hline \hline
  $d_{\rm c} (h^{-1}\kpc)$ & $10$ & $20$ & $30$ & $40$ & \\
  \hline
  $\langle A_{\rm p}(d_{\rm c})\rangle$ & 0.92 & 0.79 & 0.79 & 0.86 & \\
  \hline \hline
  $d_{\rm c} (h^{-1}\kpc)$ & $50$ & $100$ & $150$ & $200$ & $300$\\
  \hline
  $\langle A_{\rm p}(d_{\rm c})\rangle$ & 0.95 & 0.90 & 0.87 & 0.84 & 0.73 \\
  \hline \hline
\end{tabular}
\end{center}
\end{table}

By combining Equations (\ref{eq:offset_ratio}), (\ref{eq:specificP}), and
(\ref{eq:project_factor}), we have
\begin{equation}
  \langle S_{\rm p}(d_{\rm phy}>d_{\rm c}|M_{0},\ \xi,\ P,\ V)\rangle=\langle
A_{\rm p}(d_{\rm c})\rangle \cdot  \kappa^{-1}R(d>d_{\rm c}).
\label{eq:Spmean}
\end{equation}

\subsubsection{The merger rate $B$} \label{sec:prob_rate_mergerrate}

We approximate the merger rate of cluster pairs by using the following
universal fitting form of the halo merger rate obtained in \citet[][see also
\citealt{Genel2009}]{Fakhouri2010}:
\begin{eqnarray}
  B(M_0,\ \xi,\ z) & = & A_B
\cdot\left(\frac{M_0}{\tilde{M}}\right)^{\alpha}\cdot\xi^{-\beta}\cdot
\exp\left[\left(\xi \cdot \tilde{\xi}\right)^{-\gamma}\right]\nonumber \\ & &
\cdot(1+z)^{\eta}\cdot n(M_0,\ z),
 \label{eq:merger_rate} \end{eqnarray}
where $A_B=0.0104$, $\tilde{M}=1.2\times10^{12}\msun$, $\alpha=0.133$,
$\beta=-1.995$, $\tilde{\xi}=0.00972$, $\gamma=0.263$, $\eta=0.0993$. Note that
the mass ratio defined in this work is the reciprocal of that in
\citet{Fakhouri2010}. For the dark matter mass function $n(M_0,\ z)$, we use
the universal form derived from the N-body simulations (\citealt{Tinker2008};
see also \citealt{ST1999, Jenkins2001}),
\begin{equation}
n(M_0,\ z)=f(\sigma)\frac{\overline{\rho}_m}{M_0}\frac{d\ln\sigma^{-1}}{dM_0},
\end{equation}
where $\overline{\rho}_m$ is the present mean mass density of the universe,
$\sigma$ is the square root of the variance of mass, and $f(\sigma)$ gives the
fraction of the mass associated with halos in a unit range of $\ln\sigma^{-1}$
(see eq.~3 in \citealt{Tinker2008}).

\subsubsection{Initial relative velocity distribution $f_V(V)$}
\label{sec:prob_rate_v}

We approximate the probability distribution function of the pairwise velocity
obtained from cosmological simulations with halo masses above $10^{14}\msun$
\citep{Thompson2012} as the distribution of initial relative velocity $f_V(V)$
in this work. The $2016 h^{-1}\Mpc$ box size employed in the simulation of
\citet{Thompson2012} guarantees our requirement for the statistic analysis of
massive major merging systems. The best-fit of a skewed normal distribution to
the simulation results obtained in \citet{Thompson2012} is
\begin{eqnarray}
  f_x(x)& = &
\frac{1}{\sqrt{2\pi}w}\cdot\exp\left[-\frac{1}{2}\left(\frac{x-e}{w}\right)^2\right]
\nonumber \\
& & \cdot\left[1+{\rm erf}\left(\frac{a}{\sqrt{2}}\cdot
\frac{x-e}{w}\right)\right], \label{eq:f(v)}
\end{eqnarray}
where $x=\log_{10}(V/ \kms)$, $\int f_x(x)dx=1$, and the best fit parameters
are $(a,\ e,\ w)=(-2.19,\ 2.90,\ 0.295)$. The peak of this distribution locates
at 500--600 $\kms$. However, the velocity distribution could depend on cluster
masses, which was not considered in \citet{Thompson2012}.
Some other works in the literature also discussed the
relative motion of cluster pairs. For example, \citet{Dolag2013} presented the
dependence of the relative motions of cluster pairs on their distance, where
the median relative velocity can be described by a simple functional form of
$\langle V\rangle=270 \kms+1000 \kms\times d^{-1}$ (here $d$ represents the
separation of the two clusters and is measured in units of the sum of their
virial radii). In our simulations, the initial separation of two merging
clusters is set to $d=2$, and thus correspondingly the median relative velocity
is $\langle V\rangle=770\kms$ if adopting the results from \citet{Dolag2013},
which is about $25\%$ larger than the median value of Equation (\ref{eq:f(v)}).
In addition, \citet{Wetzel2011} suggested that the average value of the
infalling velocities of the satellite halos shown in cosmological simulations
is approximately the circular velocity of the primary halo at the virial radius
($v_{\rm c}$), which is smaller than that adopted in our study.

Based on the relative velocity distribution obtained from \citet{Thompson2012},
we estimate the probability of the flyby mode defined in
Section~\ref{sec:result_map_v}. It is smaller than $1\%$ of the whole merging
events, which is the reason why we only use the merger rate $B(M_{1},\ \xi,\
z)$ in Equation (\ref{eq:N_offset}) but ignore the contribution from the rare
flyby mode. We ignore the redshift evolution of the pairwise velocity
distribution for clusters, which seems insignificant at redshift $z\sim$ 0--0.3
\citep{Thompson2012, Dolag2013, Watson2014}.

\subsubsection{The impact parameter distribution $f_P(P)$}
\label{sec:prob_rate_p}

The distribution of impact parameters for major mergers of massive clusters was
investigated in the literature, and it is not easy to give a universal
quantitative description for this distribution. We list some works below.
\citet{Sarazin2001} suggested that most mergers are expected to involve fairly
small impact parameters comparable to the sizes of the gas cores in clusters,
which may be also biased to a lower value if most mergers occur along
large-scale structure filaments. \citet{Wetzel2011} (see also
\citealt{Vitvitska2002, Benson2005}) presented the distributions of the radial
($v_{\rm r}$) and the tangential ($v_{\rm t}$) velocities of DM substructures
at the time of crossing within the virial radius of a larger host halo, where
the peak of the distribution is centered on $v_{\rm r}\approx0.89v_{\rm c}$ and
$v_{\rm t}\approx0.64v_{\rm c}$ and the impact parameters are implied to be a
few hundred kpc. However, \citet{Vitvitska2002} reported that the tangential
velocity decreases with the increase of the secondary mass, and major mergers
are significantly more radial than minor mergers. In addition,
\citet{Benson2005} found the evidence for a mass dependence of the
distributions of orbital parameters, i.e., the orbits of more massive merging
systems are more radial and less tangential, though their small sample size
limits an accurate determination of this dependence.  According to the argument
in \citet{Poole2006}, the statistics including the secondary or tertiary
encounters with the primary one tends to overestimate the tangential velocity,
and the average impact parameter is likely to be smaller than what $v_{\rm
t}/v_{\rm r}$ implies in \citet{Benson2005}.  \citet{Khochfar2006} also
reported a distribution of the impact parameters, which shows a peak around
$350 h^{-1}\kpc$; however, the application of that result to the analysis of
major mergers of galaxy clusters would be limited as the simulation box size is
not sufficiently large enough.

In this work, we construct the distribution of the impact parameters by using
the following form,
\begin{equation} \label{eq:f(p)}
f_P(P)=A_P \cdot
\left(\frac{P}{\lambda}\right)^{\mu}\cdot\exp\left(-\mu\frac{P}{\lambda}\right),
\end{equation}
where $\lambda,\ \mu$ are the free parameters to control the position and the
width of the distribution peak, and $A_P$ is the normalization. When the
parameters are $(A_P,\ \lambda,\ \mu)=(7.620\times10^{-3} h\kpc^{-1},\ 327.1 h^{-1}\kpc,\ 1.636)$, Equation (\ref{eq:f(p)}) reduces to the best-fit distribution (eq.~11) obtained
in \citet{Khochfar2006}. In this work, we use $(A_P,\ \lambda,\ \mu)=(5.0\times10^{-3} h\kpc^{-1},\ 200.0 h^{-1}\kpc,\ 1.0)$ as the fiducial model, which gives the distribution peak at $200 h^{-1}\kpc$.  In Section \ref{sec:prob_prob} below,
we also try various choices of the parameters to test the dependence of the
probability of the SZ-X-ray offsets on the distribution of the impact
parameters.

\subsection{Results of the expected offset probability}
\label{sec:prob_prob}

We use Equation (\ref{eq:N_offset}) to estimate the offset rate. In our
calculation, the integration limits in Equation (\ref{eq:N_offset}) are set as
follows. Current SZ surveys have discovered clusters in a large mass range from
a few $10^{14}$ to a few $10^{15}\msun$ \citep{Williamson2011,Reichardt2013}.
We set $M_{\rm min}=2\times10^{14} h^{-1}\msun$ and $M_{\rm max}=3\times10^{15}
h^{-1}\msun$ in our calculation. We also investigate the results with different
$M_{\min}$ in Figure~\ref{pic:test_M}.  We set $\xi_{\min}=1$ and
$\xi_{\max}=3$, as only major mergers could produce the obvious offsets in the
simulations. We set $V_{\max}=V_{\rm crit}$ (see Eq.~\ref{eq:V_c}), as we
ignore the flyby mode in this work; and $V_{\rm min}$ is $10 \kms$.  We set
$P_{\min}=0$ and $P_{\max}=600 h^{-1}\kpc$. We test for a larger value of
$P_{\max}$ and find no much difference in the result, as mergers with
$P>400\kpc$ only induce small offset ratio.

\begin{figure}%[H]
  \centering
  \includegraphics[scale=0.8]{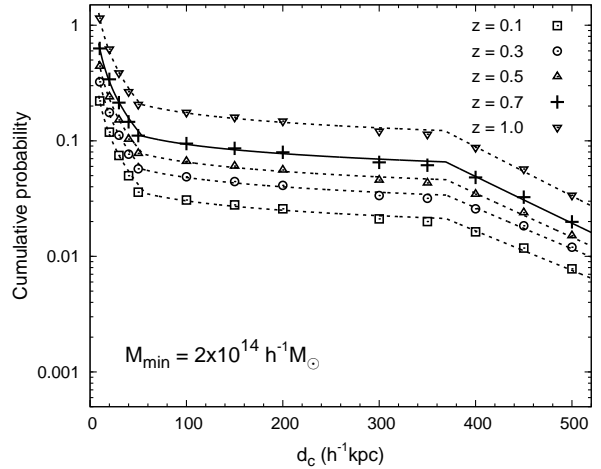}
\caption{Cumulative probabilities of the clusters with offsets larger than
$d_{\rm c}$ at different redshift, from 0.1 to 1.0. The best fits to the points
are shown by the solid ($z=0.7$, i.e., Eqs.~\ref{eq:fit_cumu_2} and
\ref{eq:fit_cumu_1}) and dotted lines. As seen from the figure, the higher
redshift case shows a larger cumulative probability of $d_{\rm c}>50
h^{-1}\kpc$. See Section~\ref{sec:prob_prob}. }
\label{pic:test_Z}
\end{figure}

We define the cumulative offset probability of the SZ-X-ray offset as follows,
\begin{eqnarray} \label{eq:cumu_prob}
& &   \mathbb{P}_{\rm cumul}(d_{\rm phy}>d_{\rm c}|z) \nonumber \\
&  = & \frac{dN(d_{\rm phy}>d_{\rm c}|z)/dz\cdot |dz/dt|\cdot (t_{\rm
relax}-t_{\rm 1st})}{\int^{M_{\rm max}}_{M_{\rm min}} n(M,\ z)dM}.
\end{eqnarray}
In Figure~\ref{pic:test_Z}, we show the cumulative probability as a function of
$d_{\rm c}$ at different redshifts. As seen from the figure, the cumulative
probability is smaller while $d_{\rm c}>50 h^{-1}\kpc$ at lower redshift.  The
probability for significant offsets observed at higher redshift is higher
because the factor of $|dz/dt|(t_{\rm relax}-t_{\rm 1st})$ in Equation
(\ref{eq:cumu_prob}) is larger at higher redshift.
For all redshift cases, the cumulative probability is flat over the range of
$50<d_{\rm c}<300 h^{-1}\kpc$. This is consistent with the `jump effect' shown
in Section~\ref{sec:result} that the offset rapidly increases to a few hundred
kpc after the small cluster passes through the larger one, which roughly equals
to the displacement between the mass density centers of the two clusters.
\citet{Forero-Romero2010} investigated the distribution of displacements
between the peaks of the DM and the gas mass densities in clusters from a large
non-radiative SPH $\rm \Lambda CDM$ cosmological simulation. They found that
about $10\%$ of the massive clusters (with masses ranging from
$2.0\times10^{14} h^{-1}\msun$ to $2.5\times10^{15} h^{-1}\msun$) at redshift
$z=0.5$ have displacements larger than $100\kpc$, and about $3\%$ are larger
than $200\kpc$. Their results are roughly consistent with ours for $d_{\rm
c}\sim$50--100$h^{-1}\kpc$; however, when $d_{\rm c}>100 h^{-1}\kpc$, their
probability is several times smaller than our results. The difference is caused
by the jump effect of the X-ray position as discussed above. The period when
the X-ray position locates around the center of the small cluster is commonly
$0.5\Gyr$ or longer. This significantly enhances the probability of the offset
larger than $100 h^{-1}\kpc$, which however might be omitted by the used
hierarchical friends-of-friends algorithm.

We calculate the cumulative probability for $d_{\rm c}>300 h^{-1}\kpc$ by
extrapolating Equation (\ref{eq:fit_R1}). As shown in Figure~\ref{pic:test_Z},
the probabilities reveal an obvious decay when $d_{\rm c}\ga 350 h^{-1}\kpc$,
mostly because $M_{\rm min}$ in Equation (\ref{eq:N_offset}) is larger when $V$
is close to the peak of the velocity distribution (see Eq.~\ref{eq:d_max}).

Specifically, we discuss the probability at $z=0.7$ in detail for two reasons, (1) the median redshift of the SZ sample in observations is nearly 0.5 \citep{Marriage2011,Reichardt2013}; (2) the average redshift of the observed SZ clusters to be compared with our estimation is $0.7$ (see Section \ref{sec:observation}). As seen from Figure~\ref{pic:test_Z}, for the significant offsets with $d_{\rm
c}=50,\ 100,\ 150,\ 200,\ 300 h^{-1}\kpc$, the cumulative probabilities of the
unrelaxed clusters with SZ-X-ray offset larger than $d_{\rm c}$ are 11.1\%,
9.5\%, 8.6\%, 8.0\%, 6.5\%, respectively. The best fit to these probabilities
by a power-law form ($\propto d_{\rm c}^{b_{\rm l}}$) is shown as the dotted
line, with $b_{\rm l}=-0.27$. As discussed above that significant offsets might be omitted in \citet{Forero-Romero2010}, our
best-fit power index ($b_{\rm l}=-0.27$) is flatter than their result ($b_{\rm l}=-1.0$).

In addition, the probability of $d_{\rm
c}=500 h^{-1}\kpc$ is approximately one third of that of $d_{\rm c}=300
h^{-1}\kpc$ at $z=0.7$. We assume an exponential decay to describe this behavior and to
extend the best-fit power-law form. Consequently, we fit the cumulative probability for $d_{\rm c}>50 h^{-1}\kpc$ as follows,
\begin{eqnarray} \label{eq:fit_cumu_2}
  & &  \mathbb{P}_{\rm cumul}(d_{\rm phy}>d_{\rm c})  \nonumber \\ & = &
     \left\{
  \begin{aligned}
     & a_{\rm l}\left(\frac{d_{\rm c}}{d_{\ast}}\right)^{b_{\rm l}}, {\ \rm if
\ }50 h^{-1}\kpc\leq d_{\rm c} \leq x_d,\\
     & a_{\rm l}\left(\frac{d_{\rm c}}{d_{\ast}}\right)^{b_{\rm
l}}\exp\left(\frac{x_d - d_{\rm c}}{x_{\rm s}}\right), {\ \rm if\ } d_{\rm
c}>x_d,
  \end{aligned}
     \right.
\end{eqnarray}
where $a_{\rm l}=0.085,\ b_{\rm l}=-0.27$ are the best-fit parameters, and
$d_{\ast}$ is set to $140 h^{-1}\kpc$. The best fits of $x_d$ and $x_{\rm s}$
are $369$ and $115 h^{-1}\kpc$, respectively.

For the offsets with $d_{\rm c}<50 h^{-1}\kpc$, the jump effect has little
influence on the cumulative probability. Our results show that the
probabilities at $z=0.7$ are $62.9\%,\ 34.0\%,\ 21.4\%$, and $14.6\%$, when
$d_{\rm c}=10,\ 20,\ 30$, and $40 h^{-1}\kpc$, respectively. We follow equation
(1) in \citet{Forero-Romero2010} to fit the results as
\begin{eqnarray}
   \mathbb{P}_{\rm cumul}(d_{\rm phy}>d_{\rm c})& = & a_{\rm
s}\left(\frac{d_{\rm c}}{d_{\ast}}\right)^{b_{\rm s}}\exp\left(-\frac{d_{\rm
c}}{d_{\ast}}\right), \nonumber \\ & & {\ \rm if\ }d_{\rm c}<50h^{-1}\kpc.
\label{eq:fit_cumu_1}
\end{eqnarray}
The best-fit gives $a_{\rm s}=0.072,\ b_{\rm s}=-0.85$,
approximately consistent with those in \citet{Forero-Romero2010}.
Since the fitting form is divergent when $d_{\rm c}$ approaches zero, we set an
cutoff at the offset where the cumulative probability is 1.0 in Equation
(\ref{eq:fit_cumu_1}).

In Figure~\ref{pic:test_M}, we show the cumulative probabilities of the
SZ-X-ray offset with different $M_{\min}$ (Eq.~\ref{eq:cumu_prob}) in panel (a)
and the contribution of mergers in different cluster mass ranges to the
cumulative probability of those clusters with $M_{\min}=2\times10^{14} h^{-1}\msun$ in
panel (b), respectively. Mergers of the massive systems ($>5\times10^{14}
h^{-1}\msun$) tend to produce a larger offset.  The cumulative probability of
$d_{\rm c}=100 h^{-1}\kpc$ is 5\% for the $M_{\min}=1\times10^{14} h^{-1}\msun$
case, while it is about 30\% for the $M_{\min}=5\times10^{15} h^{-1}\msun$ case
(see panel a).  The results in panel (b) shows that the clusters in the low
mass range ($2\times10^{14}$--$5\times10^{14} h^{-1}\msun$) dominate the
contribution of the offset. In Table~\ref{tab:fit_cumu}, we list the best-fit
parameters of the cumulative probabilities for Equations (\ref{eq:fit_cumu_2})
and (\ref{eq:fit_cumu_1}) for different redshift and mass range (see Figs.~\ref{pic:test_Z} and \ref{pic:test_M1}).

\begin{table*} %[H]
\begin{center}
\caption{Best-fit parameters of the cumulative probabilities in Equations
(\ref{eq:fit_cumu_2}) and (\ref{eq:fit_cumu_1}) for different redshifts and
mass ranges}
 \label{tab:fit_cumu}
\begin{tabular}{lccccccc}
  \hline \hline
Fig.~\ref{pic:test_Z} &  Redshift & $a_{\rm s}$ & $b_{\rm s}$ & $a_{\rm l}$ & $b_{\rm l}$  & $x_d\ (h^{-1}\kpc)$ & $x_{\rm s}\ (h^{-1}\kpc)$ \\
  \hline
&  $z=0.1$ & 0.025 & -0.86 & 0.028 & -0.27 & 369 & 138 \\
  \hline
&  $z=0.3$ & 0.039 & -0.84 & 0.044 & -0.27 & 370 & 135 \\
  \hline
&  $z=0.5$ & 0.052 & -0.84 & 0.060 & -0.27 & 370 & 123 \\
  \hline
&  $z=0.7$ & 0.072 & -0.85 & 0.085 & -0.27 & 369 & 115 \\
  \hline
&  $z=1.0$ & 0.13 & -0.86 & 0.16 & -0.27 & 368 & 109 \\
  \hline \hline
Fig.~\ref{pic:test_M1} & Mass range ($h^{-1}\msun$) & $a_{\rm s}$ & $b_{\rm s}$ & $a_{\rm l}$ & $b_{\rm l}$  & $x_d\ (h^{-1}\kpc)$ & $x_{\rm s}\ (h^{-1}\kpc)$ \\
  \hline
&  $M_{\min}=1\times10^{14}$ & 0.026 & -1.1 & 0.040 & -0.29 & 326 & 109 \\
  \hline
&  $M_{\min}=3\times10^{14}$ & 0.13 & -0.71 & 0.14 & -0.28 & 420 & 114 \\
  \hline
&  $M_{\min}=5\times10^{14}$ & 0.27 & -0.56 & 0.26 & -0.28 & 470 & 151 \\
  \hline \hline
\end{tabular}
\end{center}
\end{table*}

The SZ surveys with large sky coverage may be able to determine both the
probabilities of the clusters with offsets larger than a given $d_{\rm c}$ and
then the offset rate $dN(d_{\rm phy}>d_{\rm c}|z)/dz$, which depends on the
cluster merger rate $B$ in Equation (\ref{eq:N_offset}) and thus help to
constrain it.

\begin{figure}%[H]
  \centering
   \subfigure[\label{pic:test_M1}]{\includegraphics[scale=0.8]{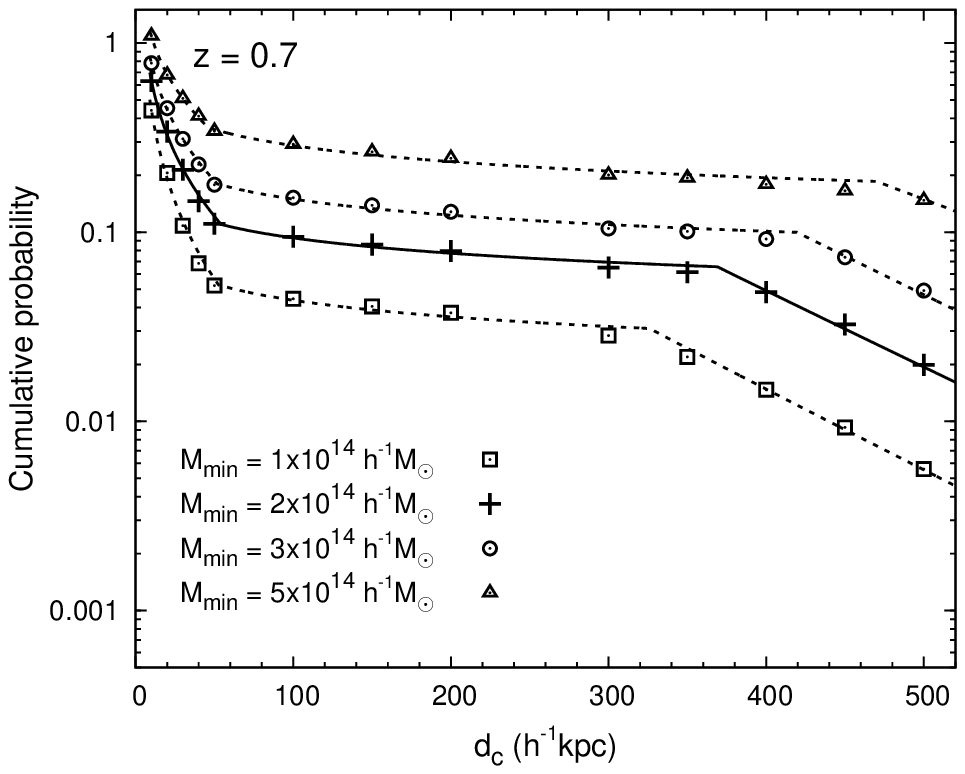}}
   \subfigure[\label{pic:test_M2}]{\includegraphics[scale=0.8]{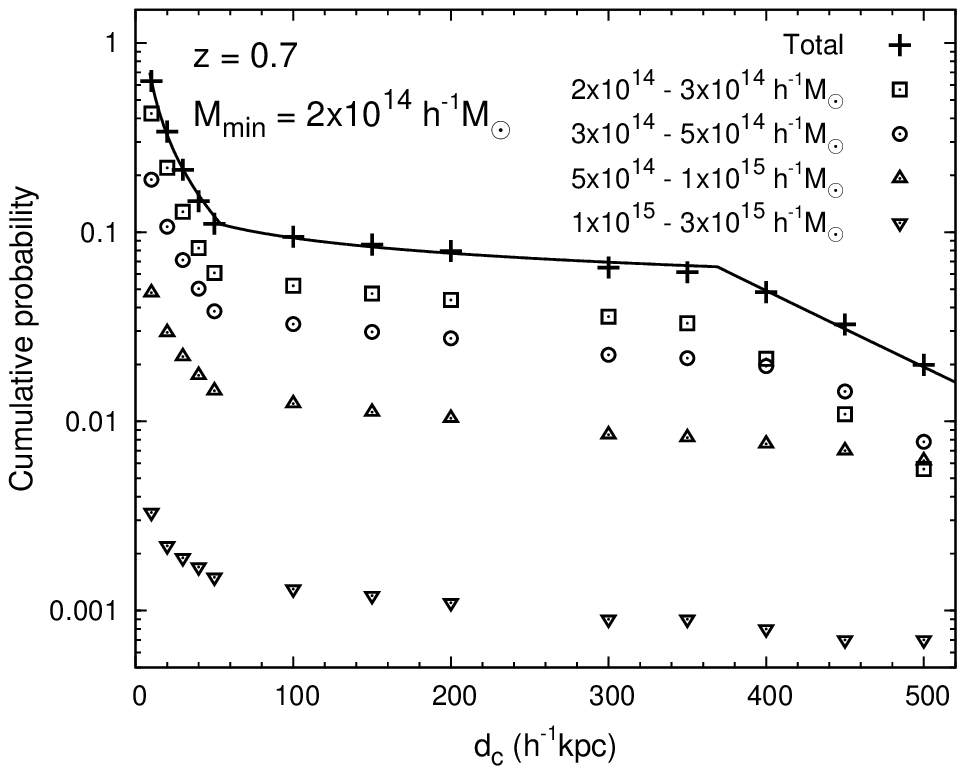}}

\caption{\textit{Top:} Cumulative probabilities for the clusters with SZ-X-ray
offsets larger than $d_{\rm c}$ at redshift $z=0.7$. The lines from top to
bottom show the results obtained by choosing different $M_{\min}$ in Equation
(\ref{eq:cumu_prob}). The cumulative probability of $d_{\rm c}>50 h^{-1}\kpc$
increases with increasing $M_{\min}$.  \textit{Bottom:} Contribution of mergers
in different cluster mass ranges to the cumulative probability obtained with
$M_{\min}=2\times10^{14} h^{-1}\msun$.  As seen from the panel, the mergers of
the clusters in the mass range $2\times10^{14}$--$5\times10^{14} h^{-1}\msun$
dominates the contribution of the offsets. }
 \label{pic:test_M}
\end{figure}

Finally, we discuss some possible uncertainties in the above estimation of the
cumulative offset probability.
\begin{itemize}
\item  In the fitting form of the offset ratio in Equations (\ref{eq:fit_R1})
and (\ref{eq:fit_R2}), we do not consider the mass ratio as an argument. The
bottom right panel of Figure~\ref{pic:offset_ratio} reveals that the offset
ratio with $\xi=3$ is about $30\%-50\%$ lower than that with $\xi=2$. Thus, we
might overestimate the probability by a factor of $\la$1.5--2. If we select a
tighter constraint of the mass ratio range in Equation (\ref{eq:N_offset}) with
$(\xi_{\min},\ \xi_{\max})=(1.5,\ 2.5)$, the cumulative probability shown in
Figure~\ref{pic:test_Z} (solid line) becomes 5.4\%, 4.6\%, 4.2\%, 3.9\%, and
3.2\% for $d_{\rm c}=$50, 100, 150, 200, and 300 $h^{-1}\kpc$, respectively,
which are nearly half of the values reported above.
\item In this work, the offset ratio $R(d>d_{\rm c})$ is assumed to be redshift
independent.  The redshift evolution shown in Figure~\ref{pic:test_Z} is purely
introduced by the halo merger rate $B(M_0,\ \xi,\ z-\Delta z)$ in Equation
(\ref{eq:N_offset}). However, the virial radius of the galaxy cluster is
proportional to $(1+z)^{-1}$; the concentration parameter is anti-correlated
with redshift; and the physical mass densities of the dark matter and gas halos
also depend on redshift. We use Figure~\ref{pic:Highz} to indicate the effect
on the maximum and the duration of the SZ-X-ray offset from the
redshift-dependent physical size of the galaxy cluster. In
Figure~\ref{pic:Highz}, we show the results of the ratio of $R(d>d_{\rm c})$ at
redshift $z=0.5,\ 1.0,\ 2.0$ to its value at redshift $z=0$, by performing a
series of cluster merger simulations with $M_1=2\times10^{14},\
5\times10^{14},\ 1\times10^{15} \msun$, $P=0 \kpc,\ V=500 \kms,\ \xi=2$.  As
seen from the figure, the deviation of the ratio from unity is smaller than
$10\%$ (or 20\%) at redshift $z=0.5$ (or $z=1$) for most of the cases. For
$d_{\rm c}\leq100 h^{-1}\kpc$, the ratios are possibly larger than $1$ at
higher redshift due to the higher central density of the cluster and thus a
strong collision intensity.  For larger $d_{\rm c}$, the ratios shift to a
lower level at high redshift.  At redshift $z=2$, the redshift evolution of the
offset ratio becomes more significant, especially for $d_{\rm c}\geq200
h^{-1}\kpc$. The maximum of the SZ-X-ray offset in the $M_1=2\times10^{14}\
M_{\odot}$ case is smaller than $200 h^{-1}\kpc$ at $z=2$, which implies that
the redshift-dependence of the cluster size ($\propto (1+z)^{-1}$) plays a more
important role on the SZ-X-ray offset size with increasing redshift, although a
higher central density enhances the strength of the collision.  We conclude
that the redshift evolution effect is an important factor for estimating the
offset probability only at $z\ga 1$.  Considering both the smaller offset ratio
$R(d>d_{\rm c})$ and the  higher mass limit of $M_{\rm min}$ in Equation
(\ref{eq:N_offset}) at high redshift, the cumulative probability of the offset
for $d_{\rm c}\geq50 h^{-1}\kpc$ may be suppressed.

\begin{figure}%[H]
  \centering
  \includegraphics[scale=0.8]{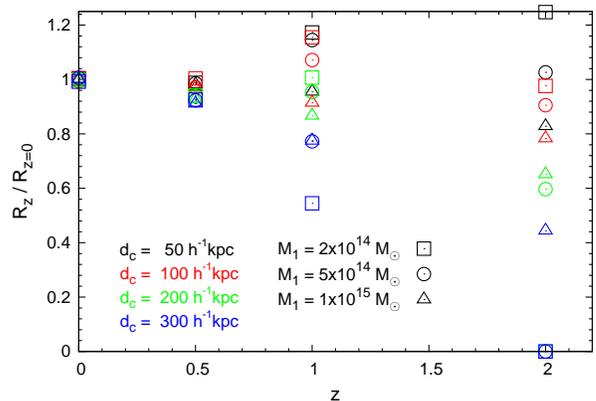}
\caption{Ratios of $R(d>d_{\rm c})$ at redshift $z$ to its value at $z=0$,
obtained from a series of cluster merger simulations with $\xi=2,\ P=0
\kpc$ and $V=500 \kms$ at redshift $z=0.5,\ 1.0,\ 2.0$. Different colors
represent the simulation results with different $d_{\rm c}$, and different
types of the points represent different cluster masses, as labeled in the
figure.  As seen from the figure, the ratios for a large $d_{\rm c}$ (e.g.,
$\ga 200 h^{-1}\kpc$) decrease at higher redshift.  }

\label{pic:Highz}
\end{figure}

\end{itemize}

\subsection{Impacts of different velocity and impact parameter distributions on the probability} \label{sec:prob_test}

In Section \ref{sec:prob_prob}, we estimate the cumulative probability of the
SZ-X-ray offset by using the fiducial model of the distributions of relative
velocities and impact parameters presented in Sections~\ref{sec:prob_rate_v}
and \ref{sec:prob_rate_p}. The forms and the parameters of the fiducial model
are motivated or obtained from cosmological simulations, but there is no
observational constraint on them so far. In this section, we discuss the
impacts of different distribution parameters on the cumulative offset
probability.

\begin{figure*}%[H]
  \centering
  \includegraphics[scale=0.9]{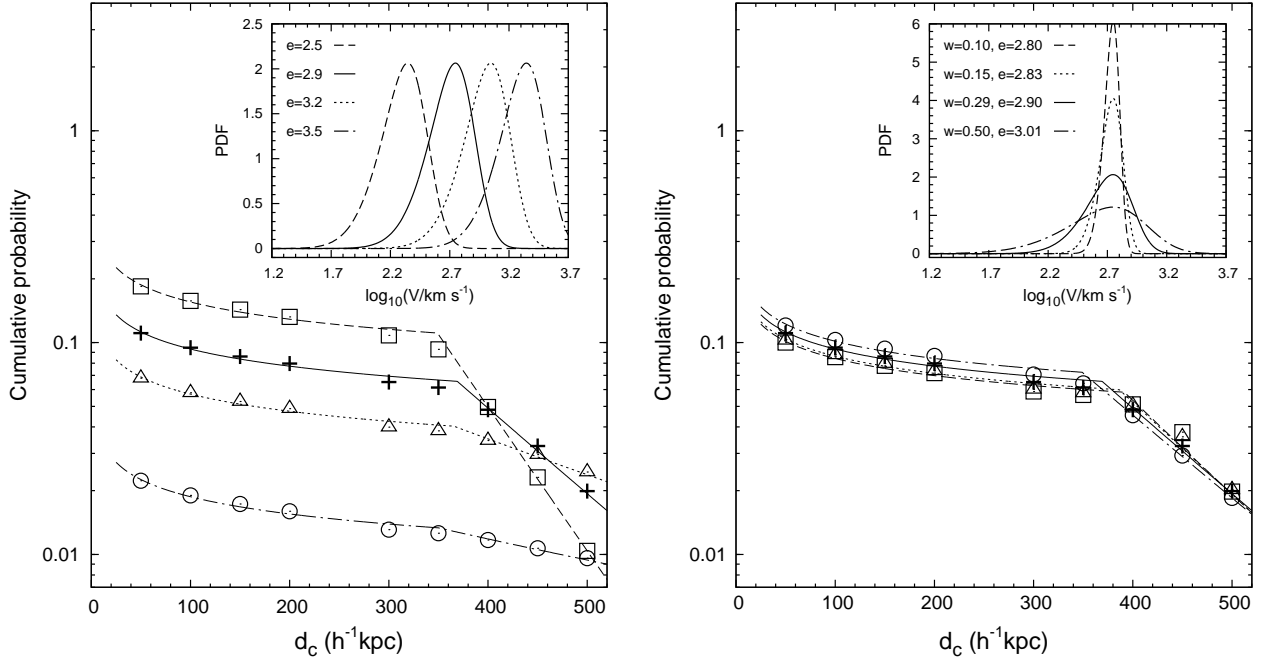}
\caption{Dependence of the cumulative probability of the SZ-X-ray offset on the
pairwise velocity distribution.  \textit{Left:} Dependence on the peak position
of the distribution. Four velocity distributions with different $e=2.5,\ 2.9,\
3.2,\ 3.5$ (see Eq.~\ref{eq:f(v)}) are shown in the inset.  Different curve
types represent different values of $e$, as labeled in the inset. The curves in
each panel are the best-fit of the probability as a function of
$d_{\rm c}$, and the points on the curves represent corresponding simulation
results. The solid line represents the fiducial model used in this paper,
similarly for the right panel. \textit{Right:} Dependence on the FWHM of the
distribution. We set $w=0.10,\ 0.15,\ 0.29,\ 0.50$ in Equation (\ref{eq:f(v)}),
and the value of $e$ is set by keeping the same peak positions as that of the
fiducial model. As seen from the figure, the cumulative probability is
sensitive to the peak position but not the width of the velocity distribution.
}
  \label{pic:test_V}
\end{figure*}

\begin{figure*}%[H]
  \centering
  \includegraphics[scale=0.9]{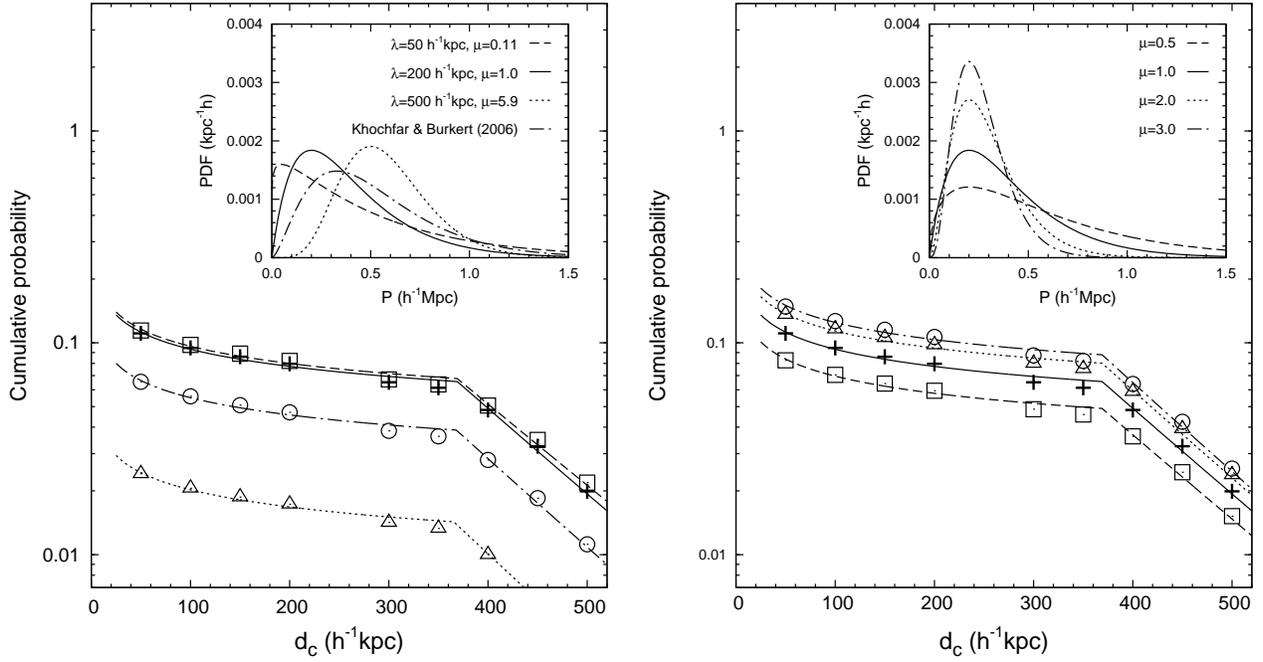}
\caption{Dependence of the cumulative probability of the SZ-X-ray offset on the
impact parameter distribution.  \textit{Left:} Dependence on the peak position
of the distribution with the same FWHM, where $\lambda=50,\ 200,\ 500
h^{-1}\kpc$ in Equation (\ref{eq:f(p)}). The probability obtained from the
distribution reported in \citet{Khochfar2006} is also presented (dot-dashed
curve). The curves have the similar meanings as those in
Figure~\ref{pic:test_V}, though for different sets of the parameters in the
impact parameter distribution.  The solid curves represent the fiducial model
used in this work. \textit{Right:} Dependence on the width of the distribution
with the same peak position $\lambda=200 h^{-1}\kpc$. The figure shows that the
amplitudes of the cumulative probabilities depend on both the peak position and
the width of the impact parameter distributions. } \label{pic:test_P}
\end{figure*}

In Figure~\ref{pic:test_V}, we explore the influence of the pairwise velocity
distribution on the probability. In the left panel, in addition to $e=2.9$ in
the fiducial model fitted from the cosmological simulation, we also obtain the
results with different values of $e$(=2.5, 3.2, 3.5) in Equation
(\ref{eq:f(v)}) to test the effects of different peak positions of the velocity
distribution, where $w$ is fixed to be $0.295$. The peaks locate in the range
of $\sim$200--2200$\kms$. In the right panel, we test the effect of different
FWHMs in the velocity distribution, where we set different values of $w$(=0.10,
0.15, 0.29, 0.50) and the value of $e$ is selected to keep the peak positions
of the distributions the same as that of the fiducial model. According to the
results shown in Figure~\ref{pic:test_V}, we find that the amplitude of the
probability is sensitive to the peak position of the velocity distribution, but
not to the width.  The probability obtained in the fiducial model is
approximately larger than that obtained with $e=3.2$ and $3.5$ by a factor of
1.6 and 5.0, respectively. As seen from the figure, the shapes of the
cumulative probabilities of $d_{\rm c}\leq300 h^{-1}\kpc$ with different
velocity distributions are approximately similar. That is because the shape of
the probability is mainly determined by the dependence of the offset ratios and
the projection factor on $d_{\rm c}$ (see Eq.~\ref{eq:Spmean}), and the offset
ratios are proportional to $d_{\rm c}^{\alpha_{\rm l}}$ in Equation
(\ref{eq:fit_R1}), where the best-fit index $\alpha_{\rm l}$ is approximately a
constant and not dependent on the parameters $V$ and $P$. The same feature also
appears in Figure~\ref{pic:test_P} for the effects of different impact
parameter distributions below. However, the cumulative probabilities of $d_{\rm
c}>300 h^{-1}\kpc$ are sensitive to the peak position of the velocity
distribution. The reason is that the mergers with high relative velocities
produce large offsets. In the $e=2.5$ case, the upper boundary of the velocity
distribution is about $600\kms$, and only the massive clusters could produce
offsets larger than $400 h^{-1}\kpc$ (see Eq.~\ref{eq:d_max}). Due to the
relatively small number of the massive clusters, the cumulative probability
decays rapidly with $d_{\rm c}>400 h^{-1}\kpc$. When the peak position of the
relative velocity distribution is higher (e.g., $e=3.2$), the cumulative
probability becomes flatter at $d_{\rm c}>300 h^{-1}\kpc$, as the relatively
less massive cluster mergers can also produce relatively large offsets.

Note that while the velocity distributions shift to the high-velocity end, the
flyby mode should not be ignored any more. From our simulations, we find that
the time durations of the significant SZ-X-ray offset of the flyby mode are
usually 1--1.5$\Gyr$ when $V\sim2000 \kms$ for different cluster masses (e.g.,
see the bottom panel of Fig.~\ref{pic:1d_v}). The durations are also inversely
proportional to the relative velocity and very sensitive to the impact
parameter (because off-axis mergers with high relative velocity, e.g., $V>2000
\kms$, are relatively ineffective in destroying the gas cores in the large
clusters, the offsets are strongly suppressed when the impact parameter gets
larger). We roughly estimate the effects of the flyby mode on our calculation.
We find that when $e=3.2$, the effect is smaller than $10\%$; and when $e=3.5$,
the probability will increase by $\sim 50\%$ after including the flyby case.
The flyby mode however does not significantly weaken the tendency that a higher
peak position of the relative velocity distribution results in a smaller
probability. Here we stress that the sensitivity of the probability to the
velocity is the crux of the Bullet Cluster problem discussed in \citet{Lee2010}
and \citet{Thompson2012}. The parameter space search of the Bullet Cluster (1E0657-56) in the literature suggested that such a system requires a high
relative velocity during the merger (e.g., \citealt{Mastropietro2008}), which
however possibly challenges the standard $\rm \Lambda CDM$ model. We suggest
that the cumulative probability of the observed SZ-X-ray offset could provide
an opportunity to examine the incompatibility existing between observed bullet
clusters and cosmological simulations.

In Figure~\ref{pic:test_P}, we show the cumulative probabilities obtained from
different distributions of the impact parameters. The left panel displays the
results of the distributions with different peak positions (i.e., $50,\ 200,\
500 h^{-1}\kpc$), but with the same FWHM. The distribution reported in
\citet{Khochfar2006} is also tested. We find that if the peak positions are
smaller than $200 h^{-1}\kpc$, the results show little difference. However, if
the peak of the distribution shifts to a larger value, the probability is
obviously suppressed as shown by the $\lambda=500 h^{-1}\kpc$ case in the
panel. Compared with our fiducial model, the cumulative probability obtained
from \citet{Khochfar2006}'s distribution is smaller by a factor 1.7.  In the
right panel, we change the width of the distribution, but keep the same peak
position. We find that a wider distribution gives smaller probability, which is
reasonable as the larger impact parameter components contribute more to the
distribution in the wider case but less to the large offsets. For the two
extreme cases shown in the panel, the probability obtained with $\mu=3.0$ is
approximately two times larger than that obtained with  $\mu=0.5$.  As a
result, we find that the amplitude of the cumulative probability ($d_{\rm c}>50
h^{-1}\kpc$) depends both on the peak position and the width of the impact
parameter distributions. However, the shape of the cumulative probability does
not, which is different from the dependence on the velocity distributions.

\subsection{The SZ-X-ray offset in observations} \label{sec:observation}

In this section, we compare our estimation of the cumulative probability of the
SZ-X-ray offset with observations in the real universe. This comparison is
motivated mainly by the following reasons: (1) in
Sections~\ref{sec:prob_prob}-\ref{sec:prob_test}, we estimate the probability
of the offset, and find that galaxy clusters with significant offsets are not
rare, for example, approximately $10\%$ of the clusters have offsets larger than $50 h^{-1}\kpc$; (2) the past several years have seen rapid progress in the SZ cluster
observation, in terms of the total numbers and the parameter ranges (precision
and redshift); and (3) the comparison would potentially provide constraints to
the model used in the estimation of the offset probability, e.g., the pairwise
velocity distribution.  The majority of the observed SZ clusters to be compared
with are at redshift $z\sim 0.7$ \citep{Andersson2011} and $1''$ corresponds to
$5 h^{-1}\kpc$ at redshift $z=0.7$.  In the following comparison, we assume
that $1''$ corresponds to $5 h^{-1}\kpc$, for simplicity.

The spatial offset between the X-ray and the SZ peaks shown in observations can
be modeled to comprise the two following components,
\begin{equation} \label{eq:d_obs}
  d_{\rm obs}=d_{\rm phy}+d_{\rm err},
\end{equation}
where $d_{\rm phy}$ is the physical one produced by the energetic merger
defined in Equation (\ref{eq:N_offset}) and $d_{\rm err}$ is the observational
error. The (differential) distribution of the physical offset $d_{\rm phy}$
follows the derivative of the offset cumulative probability (i.e.,
Eqs.~\ref{eq:fit_cumu_2} and \ref{eq:fit_cumu_1}) with respect to $d_{\rm c}$.
In this work, we assume that both of the observational errors in the spatial
positions of the X-ray and the SZ peaks follow a Gaussian distribution. The
standard deviation of $d_{\rm err}$ for the X-ray peak ($\sigma_{\rm X-ray}$)
is set to $1''$, according to the current capability of the X-ray instrument
(e.g., \textit{Chandra} X-ray Observatory). For the SZ effect, the typical position
uncertainty of the SPT SZ cluster centroid is approximately $\sigma_{\rm
SZ}=1.2'/{\rm SNR} \sim 15''$, which dominates the uncertainty $d_{\rm err}$.
In addition, we also test the results by assuming two higher resolutions of the
SZ effect ($\sigma_{\rm SZ}=2''$, and $8''$) in this work.

\begin{figure*}%[H]
  \centering
  \includegraphics[scale=0.8]{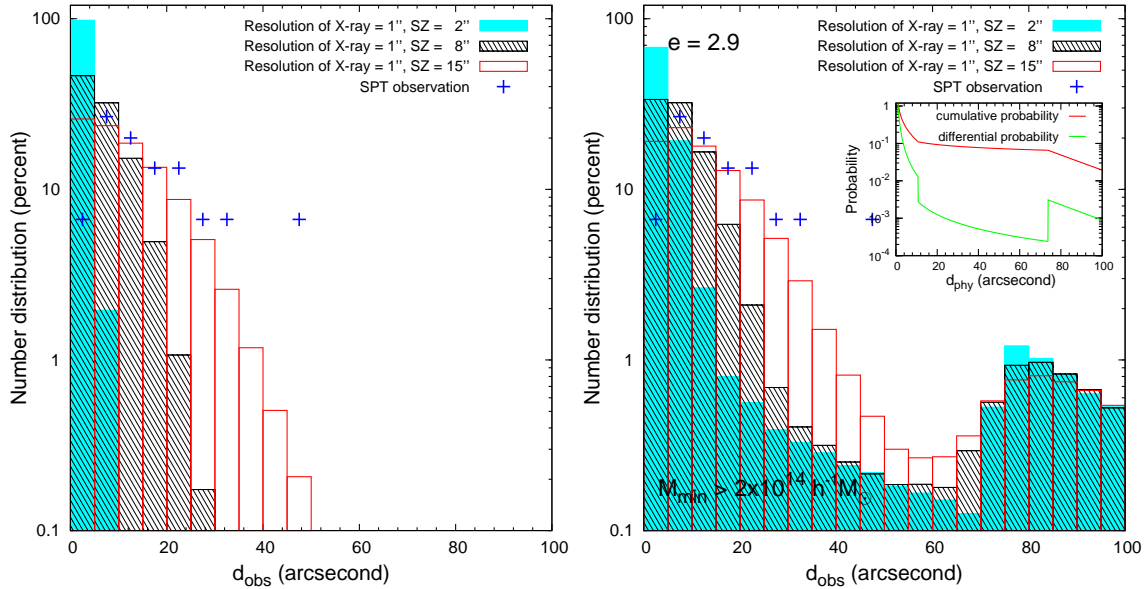}
\caption{Distributions of the offsets between X-ray and SZ peaks of galaxy
clusters. The histograms represent the results obtained from our model and
Monte-Carlo realizations, i.e., the percentage of the cluster number in each
offset bin.  \textit{Left:} The distributions obtained by assuming that the
offsets are purely contributed by observational errors (i.e., $d_{\rm phy}=0$
in Eq.~\ref{eq:d_obs}). Different histograms represent different resolutions of
SZ effects, as labeled in the panel.  \textit{Right:} The distributions
obtained by assuming that the offsets are contributed by both the physical
offsets and the observational errors.  The ``+'' points are the distributions
of the observational sample of 15 SPT SZ clusters obtained in
\citet{Andersson2011}.  The inset in the right panel displays the probability
distribution function of the SZ-X-ray physical offsets (green) and their
cumulative distribution (red). See Section~\ref{sec:observation}.
}\label{pic:resolution}
\end{figure*}

\begin{figure*}%[H]
  \centering
   \subfigure[\label{pic:reso_M}]{\includegraphics[width=0.8\textwidth]{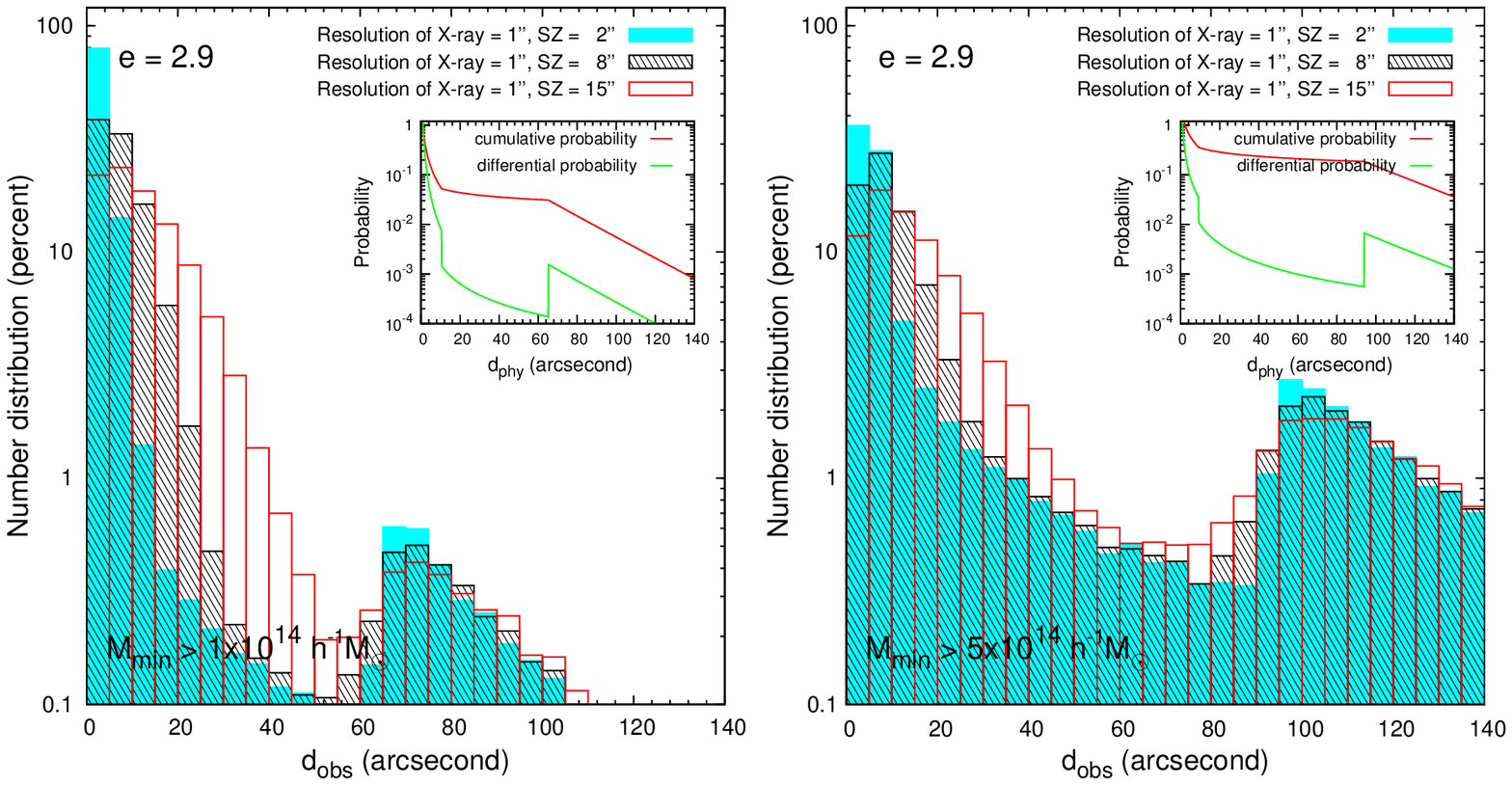}}
   \subfigure[\label{pic:reso_V}]{\includegraphics[width=0.8\textwidth]{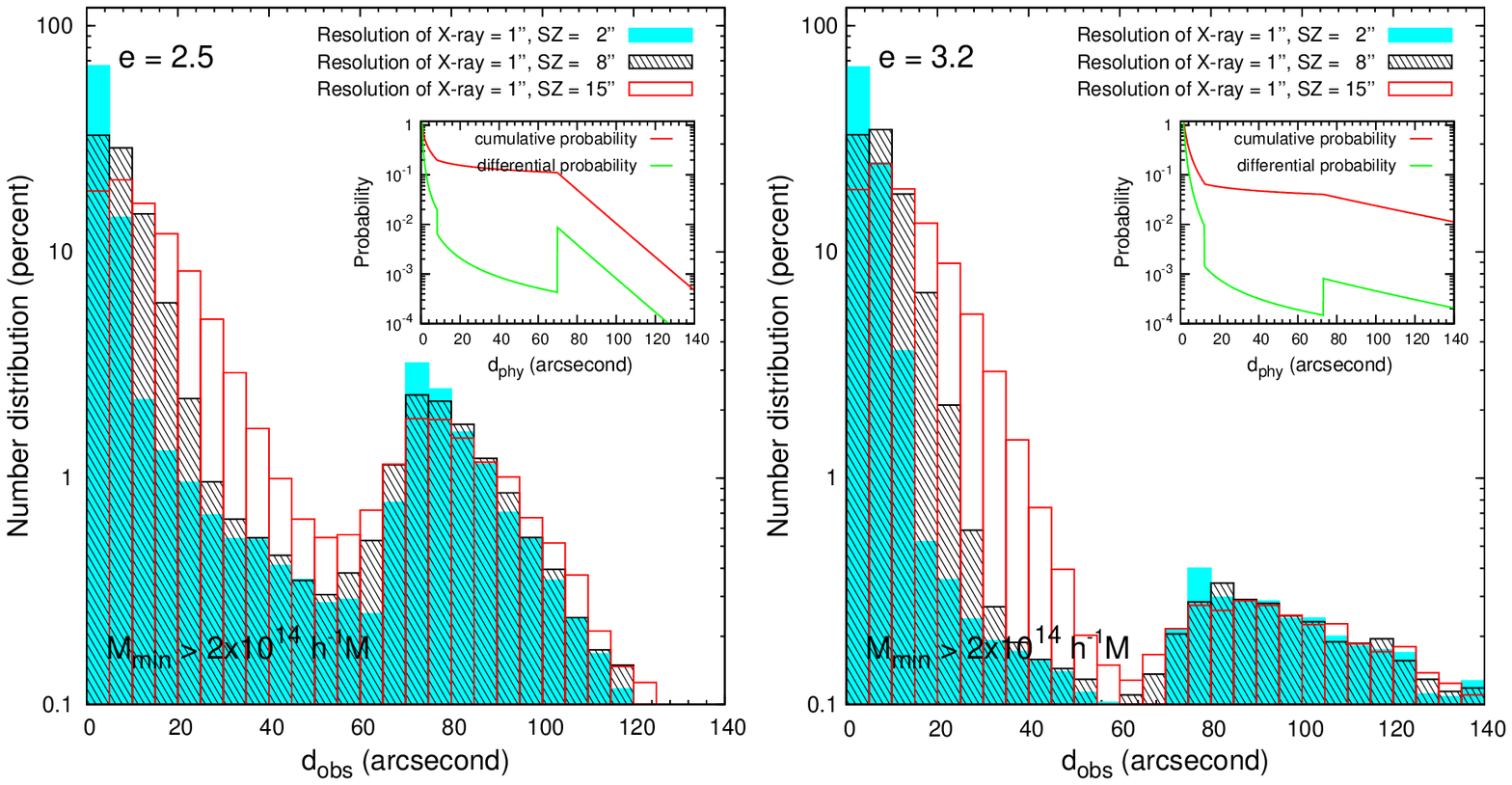}}
\caption{Distributions of the offsets between X-ray and SZ peaks of
galaxy clusters obtained with different cluster mass ranges and pairwise
velocity distributions. \textit{Top:} the distributions of the offsets with
different mass ranges $M_{\min}=1\times10^{14}$ and $5\times10^{14}
h^{-1}\msun$ (see Eq.~\ref{eq:cumu_prob}). \textit{Bottom:} the distributions
with different velocity distributions $e=2.5$ and $3.2$ (see
Eq.~\ref{eq:f(v)}). The figure shows that the bimodal distribution of the
SZ-X-ray offsets depends on the cluster mass range and the relative velocity
distribution.
} \label{pic:reso_MV}
\end{figure*}

Figure~\ref{pic:resolution} shows the statistical distribution of the SZ-X-ray
offsets $d_{\rm obs}$ by using the Monte-Carlo method to simulate both $d_{\rm
phy}$ and $d_{\rm err}$ in Equation (\ref{eq:d_obs}).  The bin size of the
offsets is $5''$ in both panels of Figure~\ref{pic:resolution}, and the
histogram represents the percentage of the number fraction of the simulated
sample in each bin.  To illustrate the effect of different distributions of
$d_{\rm phy}$, we show the observational expectations obtained by assuming that
$d_{\rm phy}=0$ and $d_{\rm phy}$ follows the derivatives of
Equations~(\ref{eq:fit_cumu_2}) and (\ref{eq:fit_cumu_1}) in the left and the
right panels, respectively. In the left panel of Figure~\ref{pic:resolution},
the offsets are contributed purely by the observational errors; while those
SZ-X-ray offsets in the right panel are contributed by both the observational
errors and the underlying physical ones.  The distribution of $d_{\rm phy}$ is
displaced in the inset of the right panel of Figure~\ref{pic:resolution} as the
solid line, and the discontinuities appear at $d_{\rm c}=10.7''$ and $73.8''$,
corresponding to the transitions of the different behavior of the cumulative
probability in three different regions of $d_{\rm c}$ as discussed in Section
\ref{sec:prob_prob} (see Eqs.~\ref{eq:fit_cumu_2} and \ref{eq:fit_cumu_1}).  As
seen from the right panel, the distribution of $d_{\rm obs}$ is bimodal once
the underlying physical distribution $d_{\rm phy}$ is considered: some clusters
peak around $d_{\rm obs}\sim 0$ and the others peak at $d_{\rm obs}\sim
70''-90''$.  The lack of clusters with the SZ-X-ray offsets around $50''-60''$
is mainly due to the ``jump effect'' as shown in Figures~\ref{pic:1d_v},
\ref{pic:1d_p}, and \ref{pic:1d_m}, respectively.  The location of the right
peak $(\sim 70''-90'')$ is determined by the combination of the following
factors: (1) the maximum offset caused by a merger increases with the cluster
mass and the initial relative velocity, as seen from Equation (\ref{eq:d_max});
(2) the cluster mass function decreases with increasing mass; and (3) the
underlying velocity distribution.  A much larger offset is more likely to be
contributed by mergers of clusters with relatively large masses and high
relative velocities (e.g., see Fig.~\ref{pic:reso_MV}). We list the predicted
SZ-X-ray offset distribution for three different SZ resolutions in
Table \ref{tab:resolution} (where the bin size is $20''$).

The distribution of $d_{\rm obs}$ may be significantly biased away from the
distribution of $d_{\rm phy}$, especially when the accuracy in determining the
SZ centroid ($\sigma_{\rm SZ}$) is not sufficient. For example, the current SPT
SZ survey ($\sigma_{\rm SZ} =15''$) may obtain mis-estimates of the SZ-X-ray
offsets $\sim 10''-20''$ for many clusters simply because of the observational
errors. However, these mis-estimates may be modelled and the underlying
physical SZ-X-ray offset distribution can still be extracted even if
$\sigma_{\rm SZ}$ is large. As seen from the right panel of
Figure~\ref{pic:resolution}, the right peak of the SZ-X-ray offset distribution
is hardly to be affected by the uncertainties in determining the SZ centroids.
If the current SZ surveys can detect many more clusters, the right peak should
be able to be revealed.  If future SZ surveys can achieve to higher resolution
and higher accuracy in determining the SZ centroids (e.g., $\sigma_{\rm
SZ}=8''$ or $2''$), the underlying physical distribution of the SZ-X-ray offset
may be better determined, and thus can be used to constrain the physics
involved in the mergers of clusters.

We show the observational offset distribution obtained from the sample of
\citet{Andersson2011} in Figure~\ref{pic:resolution} (plus signs). The
observational sample in \citet{Andersson2011} consists of 15 clusters, obtained
from observations of $178\ {\rm deg^2}$ of the sky surveyed by the SPT. The
average redshift of the sample is $z=0.68$. The observational data shown in
Figure~\ref{pic:resolution} are the values after transferring the observed
displacement between SPT detection and X-ray centroid to the distance at
redshift $z=0.7$ where $1''\sim 5 h^{-1}\kpc$.  As seen from
Figure~\ref{pic:resolution}, the uncertainties in determining the SZ cluster
centroids probably play a dominant role for those clusters with the SZ-X-ray
offsets $\la 20''$. It appears there are a few clusters with the SZ-X-ray
offsets $\ga 30''$, which cannot be due to the observational errors and must be
due to large physical offsets. Our model shows that the cumulative probability
of existing clusters with the SZ-X-ray offsets $\ga 30''-40''$ is
roughly $10\%$, which is roughly consistent with observations. Note
that the current observational sample is still small ($15$), which may lead to
large uncertainties in estimating the distribution of the SZ-X-ray offsets. If
SZ surveys can detect many more SZ clusters, it would be possible to accurately
estimate the observational SZ-X-ray offset distribution and use it to constrain
the underlying cluster merging model and the related physics involved in.  In
addition, we show that the jump effect plays a dominant role in generating
significant offsets, but here we do not consider the secondary X-ray maxima in
the data analysis, which would suppress the probability to discover the
clusters with offsets larger than $10''$.

\begin{table} %[H]
\begin{center}
\caption{Predicted SZ-X-ray offset number distributions (percentage)}
 \label{tab:resolution}
\begin{tabular}{cccccc}
  \hline \hline
  $d_{\rm obs}$ & 0--20$''$ & 20--40$''$ & 40--60$''$ & 60--80$''$ & 80--100$''$\\
  \hline
  $\sigma_{\rm SZ}=2''$ & 90.7 & 1.6 & 0.8 & 2.0 & 3.0 \\
  \hline
  $\sigma_{\rm SZ}=8''$ & 88.7 & 3.5 & 0.8 & 2.0 & 3.0\\
  \hline
  $\sigma_{\rm SZ}=15''$ & 72.9 & 18.3 & 1.9 & 2.0 & 2.8 \\
  \hline \hline
\end{tabular}
\end{center}
\end{table}

Figure~\ref{pic:reso_M} presents the model results on the SZ-X-ray offset
distribution of clusters within different mass ranges (i.e., different
$M_{\min}$ in Eq.~\ref{eq:cumu_prob}). As seen from the figure, more massive
clusters contribute larger offsets in the observation, as the right peak of the
distribution resulted from the high mass case locates at a larger spatial scale
in the high mass range (e.g., 100--$110''$ when $M_{\min}=5\times10^{14}
h^{-1}\msun$).  In Figure~\ref{pic:reso_V}, we show the distributions obtained
with different pairwise velocity distributions. As seen from the figure, the
bimodal distribution of the offsets still exist for different velocity
distributions.  As the mean value of the relative velocity increases, the `jump
effect' as shown in Figure~\ref{pic:1d_v} is more significant in causing the
scarce of the clusters with intermediate offsets. The higher relative velocity
case results in more significant offsets ($>120''$). The dependence of the
distribution of the SZ-X-ray offsets on the pairwise relative velocity
distribution of clusters suggests that the observed SZ-X-ray offset
distribution can be used to probe the cosmological velocity field at the
cluster scale.

\section{Conclusion and Discussion} \label{sec:conclusion}

In this paper, we perform a series of numerical simulations for mergers of two
galaxy clusters to understand the displacements between the spatial positions
of the maxima of X-ray and SZ maps of galaxy clusters.  The merger of two
clusters destroys their initial thermal state, and the SZ-X-ray offset is
produced due to the different dependence of the X-ray and SZ emissions on the
density and the temperature distributions of the gas.  We find significant
offsets ($\ga 100 \kpc$) mostly occur between the primary and secondary
pericentric passages of the two clusters, due to the ``jump effect''.  After
the primary core-core interaction, the densest gas region locates near the
center of the small cluster; and the X-ray peak may jump there from the center
of the larger cluster, but the SZ peak does not.

Our simulations explore the parameter space over the primary cluster masses,
cluster mass ratios, initial relative velocities and impact parameters of two
merging clusters, and we investigate the relation of the maximum and the time
duration of the SZ-X-ray offset with the simulation parameter space.  Our
findings are summarized as follows.  (1) A higher initial relative velocity
triggers a larger offset.  If the initial velocity is high enough ($\ga V_{\rm
crit}$), the two colliding clusters cannot be completely relaxed within the
Hubble time (i.e., `flyby mode'), different from the behavior of the `merger
mode'.  In the flyby mode, we find that the offset can be even up to $3\Mpc$.
However, since the pairwise velocity distribution obtained from cosmological
simulation reveals that $99\%$ of relative velocities lower than $V_{\rm
crit}$, the merger mode dominates the probability of the significantly large
offsets appearing in the Universe.  For the merger mode, the qualitative
features of the offsets caused by the mergers with initial velocity $V=500$ and
$1000 \kms$ show no significant difference.  (2) The existence and the sizes of
the offsets are sensitive to the impact parameter, since they are strongly
related with the intensity of the core-core interactions of the two clusters.
Only head-on or nearly head-on mergers can form displacements larger than $100
\kpc$, e.g., $P<400 \kpc$ for the simulation with $M_1=2\times10^{14}\msun$ and
$P<600 \kpc$ for the one with $M_1=5\times10^{14}\msun$ ($\xi=2,\ V=500 \kms$).
Mergers with smaller impact parameters result in larger sizes and longer
durations of the offsets. (3) The SZ-X-ray offsets are strongly related with
the masses of the merging clusters.  The masses of merging galaxy clusters that
possibly result in significant offsets cover the whole mass range of galaxy
clusters, i.e., $M_1>10^{14}\msun$.  (4) The mass ratio of the two merging
clusters is also an important parameter.  The significant offsets are mostly
formed by major mergers ($1<\xi<3$). For the mergers with $\xi>4$, a large
primary cluster mass $M_1>5\times10^{14}\msun$ and a high relative velocity
$V>1000 \kms$ are required to produce offsets larger than $100 \kpc$.

By applying the above results summarized from the simulations to the individual
cluster ``Bullet Cluster'' (e.g., see Fig.~10 in \citealt{Hincks2010}), we can
give a constraint on the initial relative velocity of the two merging clusters
by the SZ-X-ray offset, an observational feature different from that used in
previous work (e.g., in \citealt{Mastropietro2008}). In the Bullet Cluster, the
maximum of the X-ray image is close to the `bullet', but that of the SZ effect
locates near the centroid of the main cluster; and the displacement of the two
peaks is around $300 h^{-1}\kpc$.  Considering the high mass ratio of this
merging system \citep{Clowe2006, Bradac2006}, a relative velocity larger than
$3000 \kms$ at the initial separation $5 \Mpc$ is required to reproduce
such a significant offset in our simulations.  Note that this lower limit of
the initial velocity cannot be decreased, even if the projection effect and the
relative uncertainty in other initial parameters are considered. The constraint
on the initial velocity is in agreement with the conclusion obtained in
\citet{Mastropietro2008}, and it suggests that the SZ-X-ray offset is a good
complement to the methods used in \citet{Springel2007} and
\citet{Mastropietro2008}, where the velocity is estimated through reproducing
the morphology of the bow shock, the brightness, and the projected temperature
profile across the shock discontinuity, etc.  Constraints on the relative
velocity of the merging cluster is important, as \citet{Lee2010} (see also
\citealt{Thompson2012}) conclude that the existence of the Bullet Cluster is
incompatible with the prediction of a $\rm \Lambda CDM$ model (see also a
contrary result in \citealt{Lage2013}), unless a lower infall velocity solution
for 1E0657-56 with $\lesssim1800 \kms$ at $2R_{200}$ is found.  In our study
(see Section~\ref{sec:result_map_mr}), we show that considering the SZ-X-ray
offset and the mass ratio of the two merging clusters, there is little
possibility to find such a low velocity solution.

A high relative velocity for merging clusters like 1E0657-56 was also revealed,
for example, \citet{Molnar2012} used simulations to reproduce the SZ-X-ray
offset of the merging galaxy cluster CL J0152-1357 and found that a large
relative velocity of $4800 \kms$ is necessary to explain the observations. As
an indicator of the relative velocity, the SZ-X-ray offset owns the advantages
of easy identification, simplicity in the relation with the projection effect,
and relatively less sensitivity to the detailed gastrophysics in clusters.

To understand the statistic behavior of the SZ-X-ray offsets, we estimate the
cumulative probability of the offset, which is related with the merger rate of
galaxy clusters and the duration of the offsets in individual merging events.
We find that the cumulative probability shows different behavior depending on
whether the offset is smaller than $50 h^{-1}\kpc$ or not. This is caused by
the ``jump effect'', which could significantly enhance the probability of the
offset ($\geq50 h^{-1}\kpc$).  We also find that the mergers of the low-mass
clusters (i.e., $2\times10^{14}<M_1<5\times10^{14} h^{-1}\msun$) dominate the
contribution of the offsets for clusters larger than
$2\times10^{14}h^{-1}\msun$. The amplitude of the cumulative probability
decreases with increasing redshift.  For clusters with mass larger than
$2\times10^{14} h^{-1}\msun$ at $z=0.7$ (the average redshift of the observed
SZ clusters compared with our model results), the cumulative probabilities are
62.9\%, 34.0\%, 21.4\%, 14.6\%, 11.1\%, 9.5\%, 8.6\%, 8.0\%, 6.5\%, and 2.0\%
for SZ-X-ray offsets larger than 10, 20, 30, 40, 50, 100, 150, 200, 300, and
500$h^{-1}\kpc$, respectively. 

We further discuss some possible uncertainties in our estimation of the
probability. (1) We do not consider the mass ratio as an argument in the
fitting form of the offset ratio (see Eqs.~\ref{eq:fit_R1} and
\ref{eq:fit_R2}), which might cause an overestimation of the probability by a
factor smaller than $2$.  (2) In the simulations, we build the initial
conditions of the cluster structure at redshift $z=0$ but not consider that the
size of the cluster is redshift-dependent. We find that the redshift-dependence
has little effect on the probability when $z<1$, though the redshift effect
becomes more significant at $z>1$.

We investigate the effects of different distributions of relative velocities
$V$ and impact parameters $P$  on the probability of offset.  (1) We find that
the amplitude of the cumulative probability within $50\leq
d_{\rm c}\leq300 h^{-1}\kpc$ is partly controlled by
the relative velocity distribution of merging clusters.
For example, if the peak position of the relative velocity distribution shifts
from $550 \kms$ to $1100\kms$ (or $2200 \kms$), the obtained probability
decreases by a factor of about $1.6$ (or $5.0$).
Regarding the challenge of the Bullet Cluster to the standard $\rm \Lambda CDM$
model for which a high relative velocity is necessary, the crux of solving this
problem is the probability of high relative velocities (e.g., $>3000 \kms$),
which has been widely discussed through cosmological simulations
\citep{Hayashi2006, Lee2010, Thompson2012}. However, currently there are few
constraints from observations.  We suggest that the SZ-X-ray offset provides a
tool as the probability of the significant offsets is sensitive to the peak
position of the relative velocity distribution. (2) We find that both of the
peak position and the width of the impact parameter distribution affect the
offset probability.  If the peak position shifts from $50 h^{-1}\kpc$ to $500
h^{-1}\kpc$ with the same distribution width, the amplitude of the probability
decreases by a factor of 4.7.  As the dependence of the shape of the
probability on the impact parameter distribution is weaker than the dependence
on the relative velocity distribution at $d_{\rm c}>300 h^{-1}\kpc$, it is
robust to use the cumulative probability of the SZ-X-ray offset obtained from
observations to explore the distribution of relative velocities at the
high-velocity end.

We compare the model distribution of the SZ-X-ray offset with observations and
they are roughly consistent. However, the current sample of the SZ clusters
with the SZ-X-ray offset estimates are still small, which prevents a
comprehensive study from comparing the model distribution with the observations
and thus putting constraint on the physics involved in the merging processes of
clusters. SZ surveys, such as SPT and ACT, will detect hundreds to several
thousands of clusters. Many of them are expected to be followed up by X-ray
observations and thus have the SZ-X-ray offset measurements. With a substantial
increase of the sample size in the future, the underlying physical distribution
of the SZ-X-ray offsets can be extracted, and the unique feature of the second
peak around the $70''-90''$ and the scarce of clusters with $d_{\rm obs}$ at
$50''-60''$ may be revealed.
The physical offset distribution can be used to constrain not only the physics
involved in cluster merging processes but also the velocity field at the
cluster scale.

In this paper, we compare the cluster merging processes simulated by using the
GADGET-2 and the FLASH codes. In general, the density and temperature
distributions of the merging structure obtained from the two codes are
consistent, except that (1) the discontinuity produced by shocks are sharper in
the FLASH merger; and (2) the density of the gas core in the inner region of
the system produced by the SPH code tends to be higher (by $<5\%$). These
deviations in the strength of the surface brightness, however, have little
influence on the positions of the X-ray and the SZ peaks, which guarantees the
robustness of our simulation results obtained by merely using the SPH code.

Finally, we discuss some other possible uncertainties or assumptions in this
work. First, we do not consider the radiative cooling and various heating
mechanisms in the simulation. Among the heating mechanisms, AGN feedback is
widely proposed for the required energy in solving the overcooling problem in
clusters. However it should not be a key factor in affecting the significant
SZ-X-ray offset, because its energy is at least one order of magnitude smaller
than the gravitational binding energy released from the cluster merger. But in
the inner region of the cluster (e.g.  $\sim 100 h^{-1}\kpc$), AGN feedback has
the power to alter the baryon distribution, which reduces the central gas
density and increases the temperature in massive clusters \citep{Sijacki2006,
Sijacki2007}.  This effect should be relatively significant in the X-ray
emission since its emission is proportional to $\rho_{\rm gas}^2$, but not in
the SZ effect.  The AGN feedback might slightly increase the SZ-X-ray offset
duration $\Gamma(d>d_{\rm c})$, because the lower central gas density in the
large cluster may delay the time of X-ray peak jumping back from the center of
the small cluster. Consequently, the probability of the significant SZ-X-ray
offset will be enhanced. On the other hand, the lower gas density also mildly
reduces the collision strength of the mergers, which could decrease the sizes
of the offsets. Though the gastrophysical processes may not have significant
influence on the size of the SZ-X-ray offset as they are apt to alter the
brightness but not the position of emission peaks, more quantitative
explorations on this issue are required in the future. Second, we do not
include magnetic field in the simulation. Unless the magnetic field is
unusually strong, it should not play an important role in resulting in a
significantly large offset, because of the small energy of the normal magnetic
field relative to the mechanical energy involving in cluster mergers and its
short tangling scale relative to typical size of the SZ-X-ray offset
\citep{Carilli2002}. If the magnetic field is quite strong as simulated in
\citet{Lage2013}, the offset probability distribution around the high-SZ-X-ray
offset peak shown in this work (the right peak in Fig.~\ref{pic:resolution})
would be enhanced.  Third, our simulations show that the relaxation of the two
cluster mergers always lasts several Gyr.  In such a long relaxation time,
multiple mergers may have an influence on the SZ-X-ray offset. We run several
typical triple merger cases (the whole parameter space is too large to explore)
to examine the significance of this effect. We find that there is no apparent
difference between the size and the duration of the offsets formed by binary
and triple mergers. Though the behavior of the offset arisen by the triple
system is complicated, the possibility of the triple mergers especially both of
the two subclusters are nearly as massive as the main cluster is extremely low.
Even if multiple mergers are taken into account, the correction for the
probability of the offsets should be insignificant.

This research was supported in part by the National Natural Science Foundation
of China under nos.\ 10973001, 11273004, and 11373031. Y.L.\ is supported by
the BaiRen program from the National Astronomical Observatories, Chinese
Academy of Sciences. The software used in this work was developed in part by
the DOE NASA ASC- and NSF supported Flash Center for Computational Science at
the University of Chicago.

\end{document}